\documentclass[10pt,aps,pra,twocolumn,showpacs,superscriptaddress,nobalancelastpage,notitlepage,longbibliography]{revtex4-2}

\usepackage{graphicx,color,mathbbol,amsthm,amsmath,amsfonts,amssymb,bm,braket,footmisc,multirow,latexsym,times}

\usepackage[colorlinks]{hyperref}
\hypersetup{
 colorlinks=true,
 citecolor=blue,
 linkcolor=blue,
 urlcolor=cyan
 }

\begin{document}

\title{Nearly Heisenberg-limited noise-unbiased frequency estimation by tailored sensor design}

\author{Francisco Riberi}
\affiliation{\mbox{Department of Physics and Astronomy, Dartmouth College, 6127 Wilder Laboratory, Hanover, New Hampshire 03755, USA}} 

\author{Gerardo A. Paz-Silva} 
\affiliation{Centre for Quantum Computation and Communication Technology (Australian Research Council), \\
Centre for Quantum Dynamics, Griffith University, Brisbane, Queensland 4111, Australia}

\author{Lorenza Viola}
\affiliation{\mbox{Department of Physics and Astronomy, Dartmouth College, 6127 Wilder Laboratory, Hanover, New Hampshire 03755, USA}} 

\begin{abstract} 
We consider entanglement-assisted frequency estimation by Ramsey interferometry, in the presence of dephasing noise from general spatiotemporally correlated environments. By working in the widely employed {\em local} estimation regime, we show that even for infinite measurement statistics, noise renders standard estimators biased or ill-defined. We introduce {\em ratio estimators} which, at the cost of doubling the required resources, are insensitive to noise and retain the asymptotic precision scaling of standard ones. While ratio estimators are applicable also in the limit of Markovian noise, we focus on non-Markovian dephasing from a bosonic bath and show how knowledge about the noise spectrum may be used to maximize metrological advantage, by tailoring the sensor's geometry. Notably, Heisenberg scaling is attained up to a logarithmic prefactor by maximally entangled states, while optimal Zeno scaling is afforded by one-axis twisted spin-squeezed states. 
\end{abstract}

\date{\today}

\maketitle

\section{Introduction}

High-precision estimation of transition frequencies (or energy splittings) is a fundamental task in quantum metrology, with implications ranging from atomic spectroscopy \cite{Wineland1,Wineland2,SmerziRMP} to time-keeping with atomic clocks \cite{clocks,Ye,Colombo}. In the context of Ramsey interferometry \cite{Jon,Degen} with a quantum sensor comprising $N$ probes, the use of initial entangled states can yield, in principle, asymptotic precision bounds which surpass the optimal $N^{-1/2}$ {\em standard quantum limit} (SQL) achievable classically. Under ideal conditions, and assuming that no interaction between the probes is permitted after preparation, the ultimate $N^{-1}$ precision bound is set by the {\em Heisenberg limit} (HL), and is saturated by maximally entangled Greenberger-Horne-Zeilinger (GHZ) states \cite{Susana,Seth,Int}.

In practice, noise inevitably degrades the attainable precision, to an extent that depends on the model specifics \cite{HaaseNJP}. In many quantum sensors, dephasing noise which couples through the same operator as the signal provides the dominant noise mechanism. While no gain over the SQL can be achieved for independent Markovian noise (that is, noise with no spatial and temporal correlations) \cite{Susana,Escher}, temporal correlations alone can be exploited to achieve a superclassical {\em Zeno limit} $\propto N^{-3/4}$ \cite{Matsuzaki,Chin2012,Macies2015}. Strong temporal correlations have been detected in a variety of systems via quantum noise spectroscopy experiments, see e.g.,\,\cite{Bylander,ASuter,Romach,Marcus,Kim,Morello,walsh}, including for non-classical noise environments \cite{Quintana2017,Yan2018,Uwe}. Spatial noise correlations also tend to emerge due to probe proximity \cite{Blatt,Dorner2012,Lieven2020,Tarucha}, making noise substantially more harmful than uncorrelated one. For perfectly correlated (collective) Markovian noise, GHZ states are the most fragile, resulting in an $N$-independent precision scaling \cite{Dorner2012,DFS}, and sub-SQL scaling is also precluded in the non-Markovian regime \cite{FelixPRA,FUR}. For noise with partial spatial correlations, superclassical precision scaling can be restored by tailored error-correcting codes in the Markovian case \cite{Layden}, or by means of a randomized protocol for general temporally correlated scenarios \cite{FUR}.    

While, as the above shows, the impact noise on the scaling of precision has been extensively studied, far less attention has been devoted thus far to the fact that noise may also introduce unwanted \emph{bias}, compromising accuracy if unaccounted for. Most research on bias in quantum metrology has focused on estimation in the regime of limited data, using Bayesian approaches \cite{Haidong,Chabuda2016,Rubio}. More recently, in the context of Markovian noise, bias due to finite-frequency error-corrected sensing was addressed in \cite{CappellaroBias} through post-processing, while a purification-based protocol was proposed in \cite{Endo} to mitigate bias due to imperfect knowledge of the noise model.

Here, we focus on general {\em spatiotemporally correlated} dephasing noise, potentially arising from {\em nonclassical} sources, and address its impact on both the mean and the variance of the possible measurement outcomes. We first show how, even in the ideal limit of infinite measurement statistics and perfect knowledge of the noise spectral properties, standard estimators for both GHZ and one-axis twisted (OAT) states become systematically biased and possibly ill-defined. By suitably modifying the estimation protocol, we introduce {\em ratio estimators}, which are insensitive to dephasing by construction and match the asymptotic $N$-scaling of standard estimators. Importantly, no detailed noise knowledge is needed for constructing ratio estimators, which remain applicable in the limiting case of noise with no temporal or spatial correlations.

As our second main result, we further show that, if spatiotemporal correlations are present, access to the noise spectral features allows for the achievable precision scaling to be {\em optimized}, for a given initial state. We make our analysis concrete by examining a setting where $N$ qubits are placed on a one-dimensional (1D) regular lattice with tunable separation and couple to a bosonic bath. We show that engineering \emph{negative} noise correlations yields a far greater scaling advantage than achievable via randomized protocols that spatially decorrelate the probes on average \cite{FUR}. Remarkably, OAT states saturate the $N^{-3/4}$ Zeno limit which is optimal for independent non-Markovian dephasing \cite{Macies2015}, whereas a novel {\em nearly-Heisenberg} $N^{-1} \sqrt{\log(N)}$ scaling emerges for GHZ states.

The content is organized as follows. Section\,\ref{background} summarizes the main tools needed to understand our work, including the spatiotemporally correlated dephasing setting we consider and some basic estimation-theoretic notions used in standard metrology protocols. In Sec.\,\ref{Noisy}, we demonstrate the emergence of noise-induced bias, and show how the latter can be countered by introducing ratio estimators, at the cost of doubling the total time resources needed to carry out the experiment. The performance of these new estimators is illustrated in the simplest instance of perfectly correlated (collective) non-Markovian dephasing. In Sec.\,\ref{sec:lattice}, we specialize our analysis to a 1D lattice geometry, and show explicitly how to optimize the sensor's performance by tailoring the spatial separation of the qubit probes, while preparing them in a GHZ and OAT state, respectively. Some considerations about the robustness of our predicted superclassical precision scaling against variations of the underlying noise model are also included. We briefly summarize our conclusions in Sec.\,\ref{conclusion}, and provide complete detail about our scaling derivations for both GHZ and OAT states in a separate appendix.

\section{Background}

\label{background}
\subsection{Open quantum system setting}

We consider $N$ qubit probes, each coupled to the target frequency $b$, and a ``parallel'' dephasing noise environment (or bath). In the interaction picture with respect to the free bath Hamiltonian, $H_{\rm B}$, the joint evolution is generated by 
\begin{align}
H_{\rm SB}(t)= b J_z \otimes I_B + \frac{1}{2} \sum_{n=1}^N \,\sigma^{z}_n \otimes B_n(t)\, ,
\label{Ham}
\end{align}
where $\sigma^{u}_n$, $J_u \equiv \tfrac{1}{2} \sum_{n=1}^N \sigma_n^u$, $u \in \{x,y,z\}$, are Pauli matrices and collective spin operators, respectively, and the bath operators $\{B_n(t)\}$ describe a noise process which we take to be zero-mean, stationary, and Gaussian. Under the assumption that the initial joint state is factorized, $\rho_{\rm SB}(0)\equiv \rho_0 \otimes \rho_{\rm B}$, with $[\rho_{\rm B}, H_{\rm B}]=0$, the statistical properties of the noise are fully captured by the two-point correlation functions, 
$$C_{nm}(t) \equiv \langle B_n(t)B_m(0) \rangle =\text{Tr}_{\rm B}\{ B_n(t)B_m(0) \rho_{\rm B}\}.$$ 
For a classical noise environment, $B_n(t)$ and $\langle\bullet\rangle$ denote a stochastic commuting process and an ensemble average, respectively. In general, the noise is non-trivially correlated both in space, across different qubits, and in time (non-Markovian); ``$\delta$-correlated'' (Markovian) noise is included as a special case, $C_{nm}(t) = c_{nm} \delta(t)$, for some Hermitian matrix $c_{nm}$ that encodes the spatial noise correlations. 

In the frequency domain, the presence of temporal correlations translates into ``colored'' classical (symmetrized, $+$) and quantum (anti-symmetrized, $-$) noise spectra, given by 
$$ S^{\pm}_{nm}(\omega) \equiv \int_{- \infty}^{\infty} \!dt \,e^{- i \omega t} \langle [B_n(t),B_m(0)]_{\pm} \rangle, $$ 
with  $[B_n(t),B_m(0)]_{-}$ and $ [B_n(t),B_m(0)]_{+}$ denoting the commutator and anti-commutator, respectively, of the relevant noise operator \cite{Paz2017}. Let $| \vec{\alpha} \rangle \equiv \bigotimes_{n=1}^N |\alpha_n \rangle$, with $\alpha_n = \pm 1$ corresponding to the eigenstates $\{ |\!\!\!\uparrow\rangle,|\!\!\!\downarrow\rangle\}$ of $\sigma_n^z$, denote the $z$ basis. The time-evolved state of the system may then be generally represented as 
\begin{equation}
\langle\vec{\alpha} |\rho(t) |\vec{\beta}\rangle =e^{i b t\sum_{n=1}^N(\beta_n-\alpha_n)} e^{- \gamma(t)+ i \varphi_0(t) + i \varphi_1(t)} \langle \vec{\alpha} |\rho_0| \vec{\beta}\rangle,\nonumber
\end{equation}
 in terms of real functions $\gamma(t)$, $\varphi_0(t)$, $\varphi_1(t)$ which involve products of a state-dependent component and a corresponding time-dependent dynamic coefficient \cite{FUR}. Irrespective of the classical or quantum nature of the noise, $\gamma(t)$ governs the decay of off-diagonal coherence elements,
\begin{equation}
\gamma(t) \equiv \sum_{n,m=1}^N (\alpha_n - \beta_n) (\alpha_m - \beta_m) \kappa_{nm}(t)
\label{eq:decay}
\end{equation} 
where the decay dynamic coefficient $\kappa_{nm}(t)$ is expressible in terms of a frequency overlap integral:
\begin{eqnarray}
\kappa_{nm}(t)= \frac{1}{32\pi}\;  \!\int_{- \infty}^{\infty} \!\! d\omega\, \frac{\sin^2(\omega t/2)}{\omega^2} \,S_{nm}^+(\omega). 
\label{freqchi}
\end{eqnarray}

On the other hand, phase evolution is distinctive of a quantum, non-commuting environment.  One may show that $\varphi_0(t)$ arises from a unitary ``Lamb-shift'' contribution due to bath-mediated entanglement between the qubits, whereas $\varphi_1(t)$ is linked to whether the dephasing can be described as ``random unitary,'' hence as classical in nature \cite{FUR, Gough}. In particular, we can write 
\begin{equation}
\varphi_0(t)\equiv \sum_{n,m=1}^N (\beta_n \beta_m- \alpha_n \alpha_m) \xi_{nm}(t),
\label{phi0}
\end{equation}
with a corresponding  phase dynamic coefficient $\xi_{nm}(t)$ which can be expressed as
\begin{eqnarray}
\xi_{nm}(t)= \frac{1}{32\pi}\;  \!\int_{- \infty}^{\infty} \!\! d\omega\, \frac{\omega t -\sin(\omega t)}{\omega^2} \,S_{nm}^-(\omega), 
\label{freqphi}
\end{eqnarray}
and depends explicitly on the quantum spectra $S_{nm}^-(\omega)$. Structurally similar relationships hold for $\varphi_1(t)$ although, for symmetry reasons, this phase will not play a role in the model systems we shall focus on \cite{FUR}. Under the assumption that the spectra have vanishing support above a high-frequency cutoff, say, $S^\pm_{nm}(\omega) \approx 0$ for $|\omega| \gtrsim \omega_c$, Eq.\,\eqref{freqchi} implies a quadratic dependence upon time of $\kappa_{nm}(t)$, hence of $\gamma(t)$, in the short-time limit $\omega_c t\ll1$.  This contrasts with the linear behavior ($\dot{\gamma}(t)=\,$const) that arises in the formal limit $\omega_c\rightarrow \infty$ of Markovian noise, described by semigroup dynamics \cite{HaaseNJP}.

\subsection{Noiseless frequency estimation}
\subsubsection{Basic estimation notions}

A typical estimation protocol consists of tree distinct steps. On a first stage the $N$ qubit probes are initialized in a (possibly entangled) input state $\rho_0$. In the absence of noise, the system is then left to evolve unitarily under the Hamiltonian in  Eq.\,\eqref{Ham} with $B_n(t)=0$, for an ``encoding'' period of duration $\tau$, after which a suitable observable, say, ${\cal O}$, is measured. This process is repeated $\nu \equiv T/\tau \gg 1$ times over a total duration $T$, resulting in a vector of independently distributed measurement outcomes $\vec{\mu}\equiv \{ \mu_1, \ldots, \mu_\nu\}$, from which information about $b$ is extracted. An {\em estimator} $\hat{b}(\vec{\mu})$ is a function of random variable that associates each set of outcomes with an estimate of $b$. Let the mean and variance of $\hat{b}$ be denoted, respectively, by $\langle \hat{b} (\vec{\mu}) \rangle$ and $\Delta \hat{b}^2(\vec{\mu}) =  \langle  (\hat{b}(\vec{\mu}) - \langle \hat{b} (\vec{\mu}) \rangle )^2 \rangle$, where expectations $\langle \cdot \rangle$ are taken over all possible measurement outcomes. The estimator is {\em unbiased} if $\langle \hat{b} (\vec{\mu}) \rangle = b$ and is {\em asymptotically unbiased} if the bias vanishes as $\nu\rightarrow \infty$. More precisely, we say that $\hat{b}(\vec{\mu})$ is {\em consistent} if it converges in probability to the true value for an infinitely large sample, that is, $\forall \varepsilon >0$, $\lim_{\nu\rightarrow\infty} {\mathbb{P}} (| \hat{b} (\vec{\mu}) - b | > \varepsilon) =0$, implying $\lim_{\nu\rightarrow\infty}  \hat{b} (\vec{\mu})  =b$. A consistent estimator with vanishing variance as $\nu \rightarrow\infty$ is asymptotically unbiased. The estimators we are interested in in this work fall under this category. 

In a setting where the relationship between the expectation value of $\mathcal{O}$ and $b$ is well-known, $\langle \mathcal{O} (\tau) \rangle  \equiv f(b) $, an estimator $\hat{b}(\vec{\mu})$ is customarily constructed as follows: First, the frequency can be written as a function of  $\langle \mathcal{O} (\tau)\rangle$ by inverting the above relationship, $b\equiv f^{-1}( \langle \mathcal{O} (\tau) \rangle )$. Then, after performing $\nu \gg 1$ measurements, we estimate $\langle \mathcal{O}(\tau) \rangle$ as the sample mean of the measurements outcomes, 
$$\langle \mathcal{O}(\tau) \rangle \approx  \langle \hat{\mathcal{O}}(\tau) \rangle =\frac{1}{\nu}\, \sum_{i=1}^{\nu} \mu_i \equiv \hat{f}(b).$$ 
From this, the estimator readily follows: $\hat{b}(\vec{\mu}) =f^{-1}(\hat{f}(b))$. An analytic expression for $\Delta \hat{b}^2$ can be obtained if we further assume that the target frequency is confined within a small neighborhood of a known value $b_0$, with a corresponding mean value $\langle \mathcal{O} (\tau) \rangle_0$, such that $f(b)$ is {\em injective}. To do so, we first linearize around $b_0$,  
\begin{eqnarray*}
    \hat{b} \approx f^{-1}(f(b_0)) + \frac{\partial}{\partial f } \, f^{-1}(f(b))\big |_{f(b_0)} ( \hat{f}(b)-f(b_0)) , 
\end{eqnarray*}
whereby it follows, using the expression for the derivative of an inverse function, that 
\begin{eqnarray*}
\hat{b} - b_0  \approx  (\partial_{b}  f(\hat{b}) \big |_{b_0})^{-1}  ( \langle \hat{\mathcal{O}}(\tau) \rangle- \langle  \mathcal{O}(\tau) \rangle_0  ).
\end{eqnarray*}
Because the variance is independent of additive constants ($b_0$ and $\langle  \mathcal{O}(\tau) \rangle_0 $ in our case), it is then clear that the variance of $\hat{b} $ is proportional to $\Delta \hat{\mathcal{O}}^2(\tau)= \langle \hat{\mathcal{O}}^2(\tau) \rangle - \langle \hat{\mathcal{O}}(\tau) \rangle^2 $ in this limit, leading to the well-known error propagation formula
\begin{eqnarray}
    \Delta \hat{b}^2 (\tau) \approx (\partial_{b}\, \langle \mathcal{O} (\tau ) \rangle \big|_{b_0} )^{-2} \, \nu^{-1}\Delta \hat{\mathcal{O}}(\tau)^2.
    \label{ErrorP}
\end{eqnarray}
This procedure, known as the {\em method of moments} \cite{SmerziRMP}, has been widely used for quantifying estimation precision in metrology settings. The approximate equality in Eq.\,(\ref{ErrorP}) allows us to obtain an analytic expression for the uncertainty in the asymptotic limit of $\nu \gg 1$ and high prior knowledge, and is commonly treated as an exact measure of the frequency uncertainty. We henceforth follow this standard practice, and write the relationship in Eq.\,(\ref{ErrorP}) as an equality.  

For the particular case of an initial $N$-qubit GHZ state, $|{\rm GHZ}\rangle\equiv \tfrac{1}{\sqrt{2}}( |\!\!\uparrow\rangle^{\otimes N}\! + | \!\!\downarrow\rangle^{\otimes N} )$, we will explicitly verify that the analytic expression in Eq.\,(\ref{ErrorP}) accurately reproduces the behavior of the average over all measurement outcomes of an appropriate survival probability (see Sec.\,\ref{sub:GHZ}).  
In addition, we are also interested on frequency estimation through Ramsey interferometry for spin-squeezed initial states, which exhibit a reduced variance of the collective spin $J_u$ along a particular direction, and are easier to generate and measure experimentally than GHZ states are \cite{Nori2}. It can be shown that their uncertainty $ \Delta \hat{b} (\tau)$ can scale superclassically when measuring an angular momentum component, $\mathcal{O}= J_{\hat{n}}$. Here, we consider OAT states obtained from product coherent spin states (CSS) as follows \cite{Kita1993,Vladan2010,BraskPRX}, 
\begin{equation}
|\text{OAT} \rangle = e^{-i \beta J_x} e^{-i \theta J_z^2} |\text{CSS}\rangle_x, \qquad |\text{CSS}\rangle_x \equiv |+\rangle ^{\otimes N}, 
\label{oat}
\end{equation}
with $\beta, \theta \in [0,2\pi]$ being rotation and squeezing angles set to minimize the initial variance along $y$, and $\sigma_{x} |+\rangle = |+\rangle$. For such an OAT state, the precision $\Delta \hat{b}$ resulting from measuring $J_y$ scales like $N^{-5/6}$ in the noiseless scenario \cite{FUR}.

Lastly, we remark that if $\hat{b}$ is an unbiased estimator, the variance associated with measuring any ${\cal O}$ is lower-bounded by the {\em quantum Cram\'{e}r-Rao bound} (QCR) \cite{Helstrom,Kay}, 
\begin{eqnarray}
\Delta \hat{b}^2 \geq  \Delta \hat{b}^2_{\mathrm{QCR}} = (\nu F_{ \mathrm{Q}} [\rho_b(\tau)])^{-1},
\label{QCRB}
\end{eqnarray}
where $F_{ \mathrm{Q}} [\rho_b(\tau)]$ is the {\em quantum Fisher information} (QFI) of the output state $\rho_b(\tau)$. By construction, the QFI is derived from the maximization of its classical, operator-dependent counterpart, the \emph{Fisher Information} over the set of all positive operator valued measurements (POVMs), which includes all possible observables $\mathcal{O}$ \cite{Smerzi2014}.

\subsubsection{Case study: GHZ state}
\label{sub:GHZ}
 
\begin{figure}[t!]
\centering
\includegraphics[width=8.2cm]{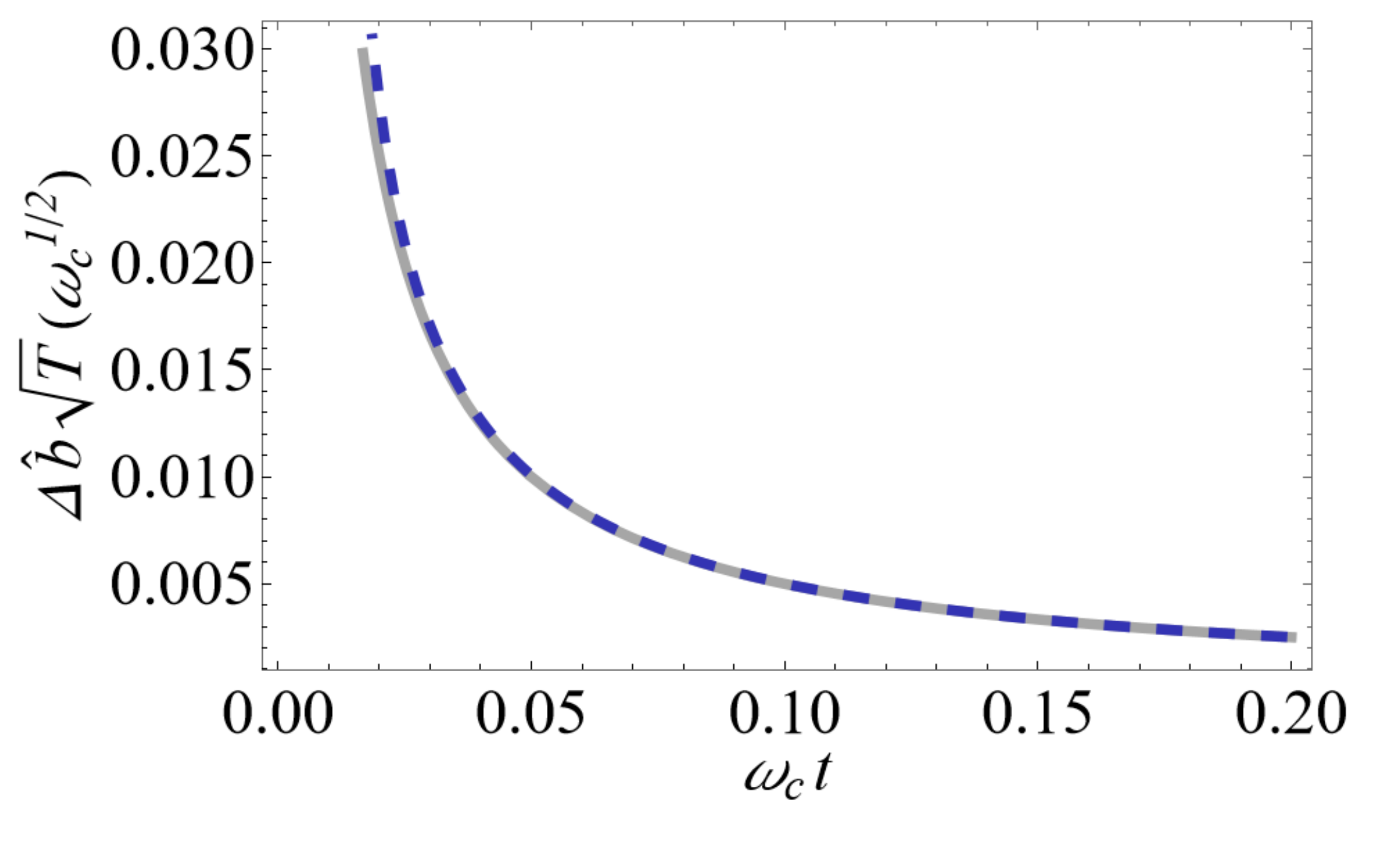}\\
\vspace*{-3mm}
\includegraphics[width=8.cm]{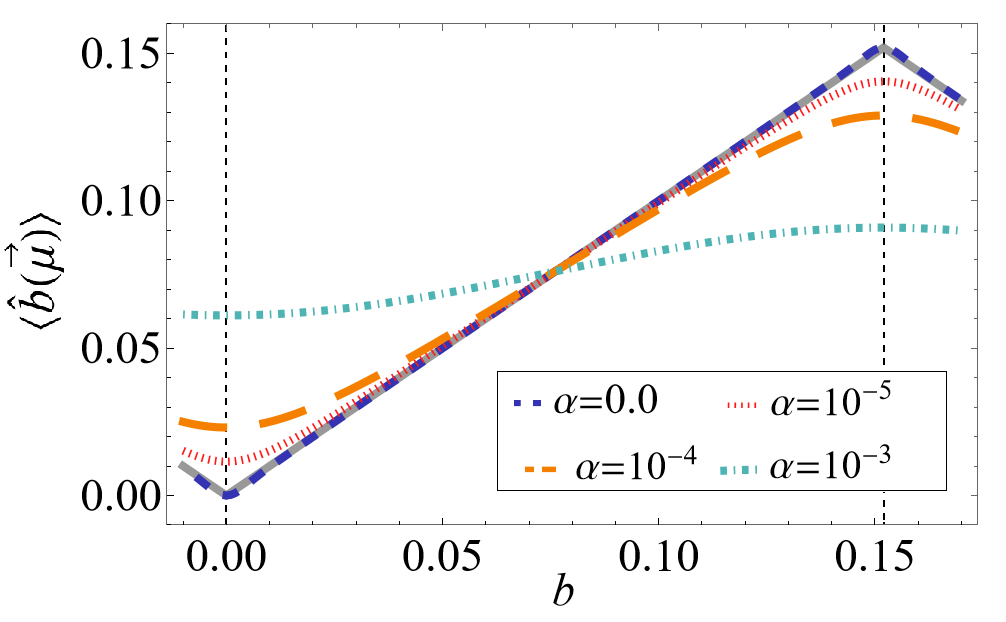}
 \vspace*{-3mm}
\caption{{\bf Standard estimator uncertainty in the noiseless setting and noise-induced bias.} Top: Standard estimator uncertainty as a function of time in the ideal limit of no dephasing. Sample variance (blue, dashed), and analytic expression  $\Delta \hat{b}(t)\approx (N \sqrt{ T t})^{-1}$ derived from the method of moments (grey, solid). Bottom: Performance of the standard estimator in the presence of {\em collective} spin-boson dephasing from a 1D zero-temperature environment (see Sec.\,\ref{SB}). The spectral density [Eq.\,\eqref{sd}] is supra-Ohmic with an exponential cutoff, with $s=3$, $\omega_c=1$,  
and different noise strengths $\alpha$, as shown in the legend. The limiting analytic expression in the absence of noise (grey, solid) is also included.  The measurement time $\tau =0.2067$ is optimal in the lattice setting. We consider an initial GHZ state with $N=100$ qubits, and $\nu =400$ measurement shots. }
\label{app}
\end{figure}

In the absence of noise, an input GHZ state saturates the HL \cite{SmerziRMP}, as can be shown by performing a measurement of the survival probability $p_{b,0}(\tau)$. Since the dynamics is unitary in this case, at a given interrogation time $\tau$ the latter is given by $p_{b,0}(\tau)=(1+ \cos(N b \tau))/2 $. Let us apply the method of moments by defining the relevant observable 
\begin{equation}
\mathcal{O}\equiv \Pi_{|\text{GHZ}\rangle}- (I -\Pi_{|\text{GHZ}\rangle}), \quad \Pi_{|\text{GHZ}\rangle}= |\text{GHZ} \rangle \langle \text{GHZ}|,
\label{o1}
\end{equation}
in terms of the projector $\Pi_{|\text{GHZ}\rangle}$ onto the initial state. In this way, $\langle \mathcal{O} (\tau) \rangle = \cos(N b \tau) \equiv f(b) $. The frequency can then be written as a function of  $\langle \mathcal{O} (\tau)\rangle$ by inverting this relationship, 
$$b=\arccos (\langle \mathcal{O} (\tau) \rangle )/(N \tau)  \equiv f^{-1}( \langle \mathcal{O} (\tau) \rangle ).$$ 
After performing $\nu \gg 1$ measurements, we can estimate $\langle \mathcal{O}(\tau) \rangle$ in terms of the ratio of the number of detections of the initial state, $\nu_+$, over the total number of trials: $ \langle \hat{\mathcal{O}}(\tau) \rangle = \hat{f}(b)= 2(\nu_+/\nu) -1 $. This leads to to the estimator 
$$\hat{b}(\tau)=f^{-1}(\hat{f}(b))= \frac{1}{N \tau} \arccos (2 (\nu_+/\nu) -1). $$ 
\noindent
Thanks to the fact that the relevant POVM has only two outcomes, the resulting measurement probability distribution is binomial, and the estimator's variance can thus be computed numerically by averaging over all measurement outcomes, 
\begin{eqnarray}
   & \Delta \hat{b}^2 (\tau)= \langle ( \hat{b}( \vec{\mu}) -\langle \hat{b}(\vec{\mu}) \rangle )^2 \rangle , 
   \label{Var}\\
    &\langle \bullet \rangle = \sum_{\nu_+=0}^{\nu}  {\nu\choose \nu_+}\; p_{b,0}(\tau)^{\nu_+}\, (1-p_{b,0}(\tau))^{\nu-\nu_+} \; (\bullet)\; \nonumber.
\end{eqnarray}
On the other hand, $\Delta \hat{b}^2(\tau)$ may be calculated by means of Eq.\,(\ref{ErrorP}). Using the second moment of the binomial distribution, $\Delta \hat{\mathcal{O}}(\tau)^2 =  \nu^{-1} 4  p_{b,0}(\tau) ( 1-p_{b,0}(\tau) ) $, Eq.\, (\ref{ErrorP}) reduces to $ \Delta \hat{b}^2 (\tau) \approx (N^2 \, T \,\tau )^{-1} $, that is, the HL. It follows that this particular choice of measurement operator $\mathcal{O}$ is optimal, as it saturates the QCR in Eq.\,(\ref{QCRB}). We can verify that such an approximate analytic expression accurately reproduces the behavior of the average over all measurement outcomes: as one may see from Fig.\,\ref{app} (top), the agreement is excellent already for $\nu =400$ measurements.

\section{Noisy frequency estimation}
 \label{Noisy}
\subsection{Noise-induced bias} 
 
Crucially, the method of moments we described above relies on the assumption that the function $f$ to be {\em one-to-one}, at least in a neighborhood of $b_0$. Despite being ubiquitous in the literature, this inversion procedure on which the construction of $\hat{b}(\vec{\mu})$ hinges upon generally becomes problematic in the presence of noise. To illustrate the issue, consider as before an initial GHZ state, but this time evolving under the dephasing Hamiltonian of Eq.\,(\ref{Ham})  which, to simplify our argument, we take to be {\em collective} (permutation-invariant), that is, $B_n(t) \equiv B(t), \forall n$.

In the presence of dephasing, the survival probability becomes  $p_b(\tau)= (1+ \cos(N b \tau)  e^{- \gamma_{\text{GHZ}}(t)})/2$  and, correspondingly, the mean value is multiplied by a decay factor, $ \langle \mathcal{O}(\tau) \rangle =  e^{- \gamma_{\text{GHZ}}(t)} \cos(N b \tau)  \equiv g(b)$. Thus, using the same inversion formula to estimate $b$ as in the noiseless scenario, $\hat{b}(\tau)=\arccos(2(\nu_+/\nu)-1)/(N \tau )= f^{-1}(g(b))$, leads to a systematic \emph{noise-induced bias} as $\gamma_{\text{GHZ}}(t)$ is not accounted for. This is illustrated in Fig.\,\ref{app}(bottom), which shows how the performance of the standard estimator increasingly degrades as the dephasing rate increases. Alternatively, computation of $\hat{b}$ using the correct inversion formula would require {\em perfect} knowledge of the decay parameter $\gamma_{\text{GHZ}}(\tau)$, 
$$\hat{b}(\tau)=\frac{1}{N\tau} \arccos\big[ e^{\gamma_{\text{GHZ}}(\tau)}(2 (\nu_+/\nu)-1)\big],$$ 
which would imply previous noise characterization to the {\em same level of precision} desired for the estimation of $b$. Not only would this add to the resource count, but there is no known noise characterization protocol capable of achieving sub-SQL scaling.  Even more importantly from a conceptual standpoint, the above recipe returns {\em imaginary} values for $\hat{b}$ whenever $e^{\gamma_{\text{GHZ}}(\tau)}(2 (\nu_+/\nu)-1)\geq 1$. Note that this happens for a non-vanishing set of outcomes whenever $\gamma_{\text{GHZ}}(\tau)>0$, rendering quantities such as $ \langle \hat{b} (\vec{\mu}) \rangle $ ill-defined. 

Despite the pitfalls involved in this procedure, the analytical expression for the uncertainty which follows from the method of moments has found widespread application in the literature. For a fixed total time $T$, we have \cite{HaaseNJP}
\begin{align}
\Delta \hat{b}(\tau)^2= \frac{1}{T \tau N^2 } \bigg[\frac{1-e^{-2 \gamma_{\text{GHZ}}(\tau)} \cos^2(N b \tau)}{e^{-2 \gamma_{\text{GHZ}}(\tau)} \sin^2(N b \tau) } \bigg].
\label{std}
\end{align}
The performance of Eq.\,(\ref{std}) can be quantitatively assessed once a specific noise model is chosen. For collective spin-boson dephasing \cite{FelixPRA}, which we will also recover as a limiting case in Sec.\,\ref{SB}, we have $\gamma_{\text{GHZ}}(\tau)= N^2 \kappa(\tau)$. Optimizing with respect to the phase argument $\varphi\equiv N \tau b$ by demanding that $\varphi= (2 k +1) \pi/2$, $k \in \mathbb{N}$, and minimizing with respect to $\tau$ in the short-time regime $\omega_c \tau \ll 1$ where 
$\gamma_{\text{GHZ}}(\tau) \approx N^2 \kappa_0^2 (\omega_c \tau)^2$ for some dimensionless constant $\kappa_0^2$, the optimal performance can be shown to be SQL-limited. Specifically, $\Delta \hat{b}^{\text{GHZ}}_{\text{opt,coll}}= e^{1/4} (\kappa_0 \omega_c/T)^{1/2} N^{-1/2}$, for an optimal measurement time $\tau^{\text{GHZ}}_{\text{opt, coll}}= 1/2 (\kappa_0 \omega_c)^{-1} N^{-1}$.

\subsection{Ratio estimator for GHZ states}
\label{RGHZ}

\begin{figure}[t!]
\centering
\includegraphics[width=8.2cm]{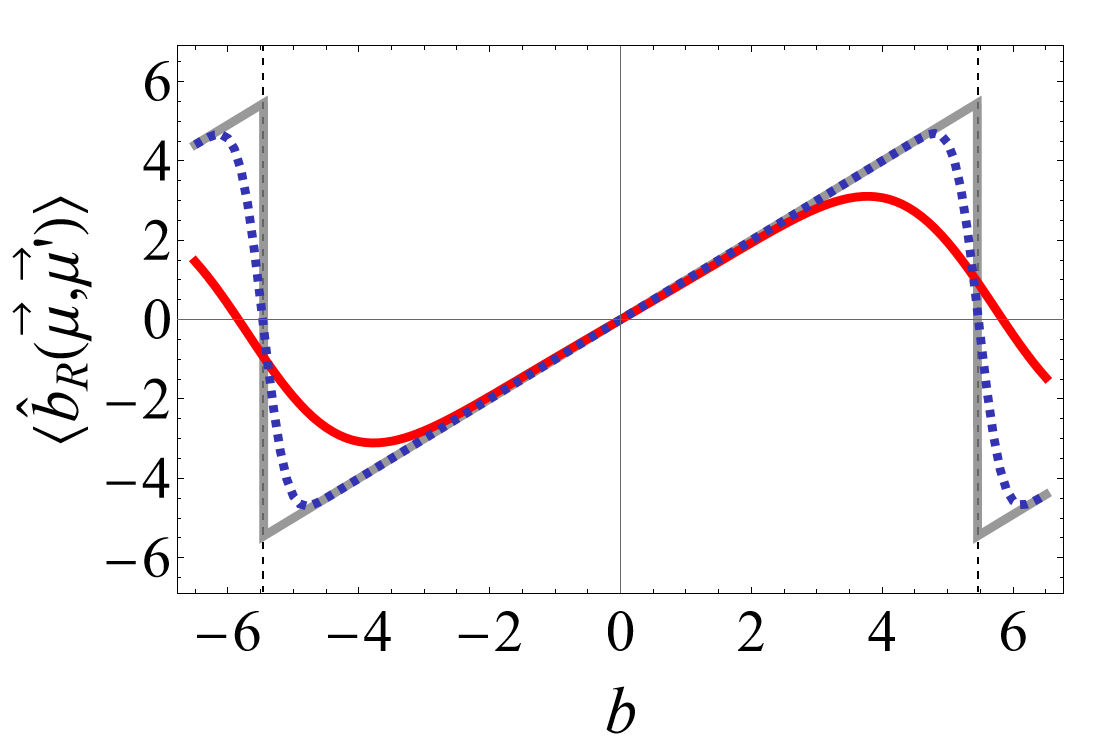} \\
\vspace*{-6mm}
\hspace*{3mm}\includegraphics[width=8.2cm]{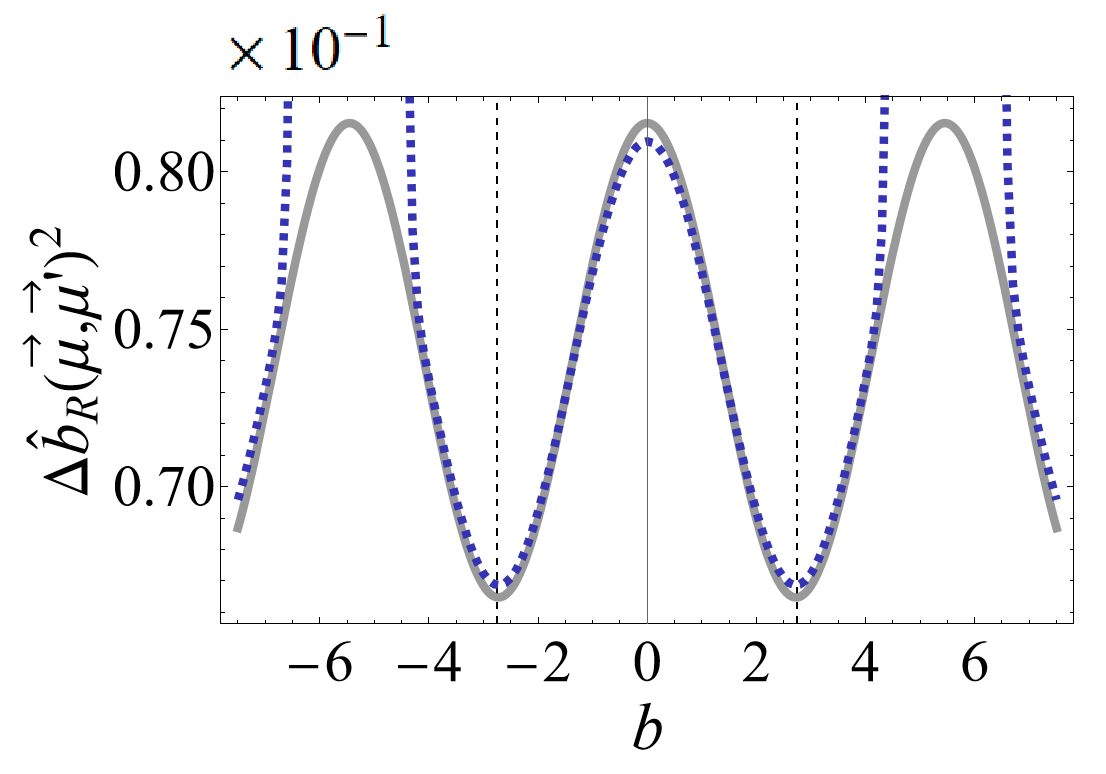}
\vspace*{-3mm}
\caption{(Color online) \textbf{Performance of ratio estimator for collective dephasing.} Top: Ratio estimator's sample mean for $\nu=30$ (red, solid), $\nu=400$ (blue, dashed), and analytic $\nu \rightarrow \infty$ mean value (grey, solid). Bottom: Ratio estimator's sample variance for $\nu=400$ and limiting analytic expression (grey, solid). In both cases, a GHZ state of $N=100$ qubits is considered, subject to  {\em collective} spin-boson dephasing from a 1D zero-temperature noise environment; a supra-Ohmic spectral density with $s=3, \omega_c=1$ and noise strength $\alpha=1$ is used, and the measurement time $\tau \approx 0.0028$ is chosen to minimize uncertainty.}
 \label{dbplot}
\end{figure}

To circumvent the inversion issues plaguing the estimator linked to Eq.\,(\ref{std}), we propose to modify the estimation protocol in such a way that {\em two} suitable observables are independently measured, and their outcomes combined to obtain a noise-robust estimator. Concretely, for the GHZ state we are focusing on, we propose to not only measure the observable $\mathcal{O}$ given in Eq.\,\eqref{o1}, but also $\mathcal{O}' \equiv  \Pi_{|{\rm{GHZ}'}\rangle}- (I -\Pi_{|{\rm{GHZ}'}\rangle}) $, where we now project into a related state within the GHZ class:
$$|{\rm{GHZ}'}\rangle\equiv  \frac{1}{\sqrt{2}} \big(|\!\!\uparrow \rangle^{\otimes N} + 
i |\!\!\downarrow \rangle^{\otimes N} \big).$$
Accordingly, the probability of measuring $\Pi_{|{\rm{GHZ}'}\rangle}$ is given by $p_b'(\tau)= (1+ \sin(N b \tau)  e^{- \gamma_{\text{GHZ}}(t)})/2$. With each observable being measured $\nu=T/\tau$ times, our protocol implies a total of $2 \nu$ measurements: that is, we {\em double} the time resources needed to carry out the experiment from $T$ to $2T$. Since $\langle \mathcal{O}' (\tau) \rangle =   e^{- \gamma_{\text{GHZ}}(t)} \sin(N b \tau) \equiv h(b) $, we can solve for the frequency by taking the ratio between the two mean values, $b= \arctan[ {\langle \mathcal{O}'(\tau) \rangle}/\langle \mathcal{O}(\tau) \rangle]/(N \tau)$. It follows that a noise-robust \emph{ratio} estimator for $b$ can be constructed as 
\begin{equation}
\hat{b}_{\text{R}}^{\text{GHZ}} (\tau) \equiv \frac{1}{N\tau} \arctan \big[  \langle \hat{\mathcal{O}}'(\tau) \rangle /\langle \hat{O}(\tau)\rangle \big] ,
\label{RE}
\end{equation}
where 
$\langle \hat{\mathcal{O}}'(\tau) \rangle=\hat{h}(b) =2 (\nu'_+/\nu) -1$ and $\langle \hat{\mathcal{O}}(\tau) \rangle=\hat{g}(b)=2(\nu_+/\nu) -1$ are the finite-sample estimators to the respective mean values, obtained from $\nu'_+$ (respectively, $\nu_+$) detections in $\nu$ trials. In Fig.\,\ref{dbplot} (top), the limiting estimator function for $\hat{b}_{\text{R}}(\tau)$, $\arctan[\tan(N b \tau)]/(N \tau)$, is compared to an average over all measurement outcomes which now involves two independent probability distributions corresponding to $\mathcal{O}$ and $\mathcal{O}'$, for two different numbers of repetitions $\nu$ and $\nu'$, showing excellent asymptotic convergence. 

In a similar way, for the variance of the ratio estimator in Eq.\,\eqref{RE}, we have 
 \begin{align}
      &\Delta \hat{b}_{\text{R}}^{\text{GHZ}}(\tau)^2= \langle ( \hat{b}_{\text{R}}( \vec{\mu}, \vec{\mu'}) -\langle \hat{b}(\vec{\mu}, \vec{\mu}') \rangle )^2 \rangle, \label{VarR}\\
      \langle \bullet \rangle &= \bigg\{\sum_{\nu_+, \nu'_+=0}^{\nu}  {\nu\choose \nu_+}\; p_{b}(\tau)^{\nu_+}\, (1-p_{b}(\tau))^{\nu-\nu_+}\nonumber\\
      & \hspace{20mm}{\nu\choose \nu'_+}\;p'_{b}(\tau)^{\nu'_+}\, (1-p'_{b}(\tau))^{\nu-\nu'_+}\; (\bullet) \bigg\}.
\nonumber
 \end{align}
Assuming that our prior knowledge confines the frequency within a neighborhood of $b_0$ in which {\em both} $g(b)$ and $h(b)$ are injective, we can linearize to obtain an analytic expression for the uncertainty in terms of the dispersions $\Delta \hat{O}^{2}(\tau)$ and $\Delta \hat{O}'^{\, 2}(\tau) = \nu^{-1} 4 \, p_b'(\tau) (1-p_b'(\tau)) $. This leads to 
\begin{eqnarray}   
   \Delta \hat{b}_{\text{R}}^{\text{GHZ}}(\tau)^2 &\!\!=\! & 
\!   \bigg(\!\frac{\partial \hat{b}_{\text{R}}}{\partial \langle \mathcal{O}'(\tau) \rangle }\!  \bigg)^2 \!\Delta \hat{\mathcal{O}}'^{2}(\tau) +\! \bigg(\!\frac{\partial \hat{b}_{\text{R}}}{\partial \langle \mathcal{O}(\tau) \rangle } \!\bigg)^2 \!
\Delta \hat{\mathcal{O}}^2(\tau) \nonumber \\
  & \!\!=\!& \frac{1}{T \tau N^2}\bigg[e^{2  \gamma_{\text{GHZ}}(\tau)}- \frac{1}{2} \sin^2(2 N b \tau) \bigg],
   \label{dbGHZ}
\end{eqnarray} 
to be contrasted with Eq.\,(\ref{std}). Interestingly, in this case the optimal performance is achieved when the phase $\varphi'\equiv 2 N b\tau$ is $\varphi =(2k +1) \pi/4$. In Fig.\,\ref{dbplot} (bottom), the analytic expression $\Delta \hat{b}_{\text{R}}^2(\tau)$ in Eq.\,(\ref{dbGHZ}) is compared against the sample variance for finite $\nu$, showing remarkable agreement in the region where the estimator is linear. Note that the sample variance diverges around the points where the estimator is discontinuous, forcing our prior knowledge of the frequency to be confined to an interval of length $\pi/(N \tau)$.

In order to compare the performance of $\Delta \hat{b}_{\text{R}}(\tau)^2$ against the standard figure of merit of Eq.\,(\ref{std}), for consistency it is necessary to let $T \mapsto 2T$ in the latter, so that the same fixed resources are ensured. It can be checked, by plotting both uncertainties as a function of time, that $\Delta \hat{b}_{\text{R}}(\tau)$ is slightly bigger than its standard counterpart at all times (see inset of Fig.\,\ref{app2}(left)). The optimal interrogation time for the ratio estimator is harder to compute analytically. However, evaluating the ratio estimator uncertainty at the time $\tau^{\text{GHZ}}_{\text{opt, coll}}$ that minimizes the standard one, we can verify that it reaches the {\em same SQL scaling} as its standard counterpart, $\Delta \hat{b}_{\text{R}}(\tau^{\text{GHZ}}_{\text{opt, coll}}) = (2 \sqrt{e} -1)^{1/2} (\kappa_0 \omega_c/T)^{1/2} N^{-1/2}$, albeit with a  slightly larger constant prefactor ($(2 \sqrt{e} -1)^{1/2} \approx 1.52$ as opposed to $e^{1/4} \approx 1.28$). 

A similar analysis for the more general case of non-collective spin-boson dephasing will be carried out in the context of the noise-tailored lattice sensor of Sec.\,\ref{sec:lattice}.

\subsection{Ratio estimator for general Ramsey interferometry}

While the emergence of noise-induced bias was illustrated for GHZ states in the above discussion, standard estimators used in Ramsey interferometry with different initial states are plagued by similar issues in the presence of dephasing noise. Here, we focus on dephasing noise from a bosonic bath and a class of OAT states, as defined in Eq.\,\eqref{oat}, which includes CSS product states as a limiting case ($\theta=\beta=0$) as well as OATs with squeezing and rotation angles set to minimize the initial variance along $y$. In the $N\gg 1$ limit, we have \cite{Kita1993,FUR}: 
\begin{align}
\theta_{\text{opt}}(N) &\approx 12^{1/6}\, 2^{2/3}\, N^{-2/3} \label{ideal} \\ 
\beta_{\text{opt}}(N) &\approx \pi/2 - 3^{-1/6} N^{-1/3}- 3^{1/6} 2^{-1} \;N^{-2/3}. \nonumber
 \end{align}
After an encoding time $\tau$, the mean value $\langle J_y(\tau)\rangle$ can be shown to be given by an expression of the form (see Appendix\,\ref{expect} for more detail)
\begin{align}
\langle J_y(\tau) \rangle =  e^{- \gamma_{\text{OAT}}(\tau)} 
\mathcal{Q}(\tau) \sin(b \tau ),
\label{Jy}
\end{align}
where $\gamma_{\text{OAT}}(\tau)$ depends on decay coefficients $\kappa_{nm}(\tau)$, $\mathcal{Q}(\tau)$ is a function of both the angles $\theta, \beta$ and, for a nonclassical environment, the coefficients $\xi_{n m}(\tau)$ determine the phase evolution $\varphi_0(\tau)$.

For a product state, it is possible to obtain the exact expression $\mathcal{Q}(\tau)=\sum_n \prod_{m \neq n} \cos(  \,\xi_{n m}(\tau)) $, whereas for arbitrary squeezing and twisting angles the reduced dynamics become highly entangled and an exact analytic approach is unfeasible. However, in the short-time regime $\omega_c \tau \ll 1$ where the relevant evolution takes place, the above mean values can be computed to great accuracy by performing a second-order cumulant expansion over the qubit operators \cite{FelixPRA,FUR}. In this same limit, any quantum noise contribution due to $\varphi_0(t)$ enters as a correction to the decoherence dynamics determined by the decay terms $\kappa_{nm}(t)$. Thus, it may be disregarded to first approximation, making $\mathcal{Q}(\tau)$ approximately time-independent, say, $\mathcal{Q}_0 \equiv \mathcal{Q}_0(\theta, \beta)$. Inverting this relationship yields 
$$\hat{b}(\tau)= \frac{1}{\tau}\arcsin [ e^{ \gamma_{\text{OAT}} } \langle \hat{J}_y (\tau) \rangle \mathcal{Q}_0^{-1}  ] , $$
with $\langle \hat{J}_y (\tau) \rangle = \sum_{m=-N/2}^{N/2} (\nu_m/\nu) $ being the sample average of $\langle J_y(\tau)\rangle$ after $\nu$ detections. Similar to the GHZ case, evaluating this estimator would require precise knowledge of the noise parameters $\gamma_{\text{OAT}}(\tau)$, $\mathcal{Q}_0$, and may yield an imaginary number for the subset of outcomes where $ e^{ \gamma_{\text{OAT}}(\tau)} \langle \hat{J}_y (\tau) \rangle \mathcal{Q}_0 > 1$ in the argument of the arcsin. We now show how this issue can again be circumvented by constructing a ratio estimator after measuring two distinct observables -- in this case, the two orthogonal angular momentum components $J_x$ and $J_y$.

Since, from Eq.\,\eqref{Jy}, the target frequency $b$ enters the reduced dynamics unitarily, say, $U(\varphi)=\exp(- i \varphi J_z )$, with $\varphi = b t$, we can derive the equations of motion with respect to $\varphi$ for $\langle J^{\ell}_{v} (\varphi) \rangle$, $v \in \{x,y\}$, for the two relevant moments $\ell \in \{1,2 \}$. For $\ell=1$, taking the derivatives in $\langle J_v(\varphi)\rangle \equiv \text{Tr} \{ J_v U(\varphi) \rho_0(t) U(\varphi)^{\dagger}\}$, with $\rho_0(t)$ depending parametrically on the noisy dynamical coefficients, we obtain:
\begin{eqnarray*}
\partial_{\varphi} \langle J_x(\varphi)\rangle& = -i \langle [J_x,J_z](\varphi) \rangle = \langle J_y(\varphi) \rangle, \\
\partial_{\varphi} \langle J_y(\varphi)\rangle &= -i \langle [J_y,J_z](\varphi) \rangle = - \langle J_x(\varphi) \rangle  ,
\end{eqnarray*}
which immediately leads to $ \partial^2_{\varphi} \langle J_v(\varphi) \rangle = - \langle J_v(\varphi ) \rangle$.  Thus,
\begin{eqnarray*}
 \langle J_x (t) \rangle &= \cos(\varphi)\, \mathcal{F}(t) + \sin(\varphi)\, \mathcal{G}(t),\\
  \langle J_y (t) \rangle  &= -\sin(\varphi) \,\mathcal{F}(t) + \cos(\varphi)\, \mathcal{G}(t), 
\end{eqnarray*}
for some state-dependent functions $\mathcal{F}(t),\mathcal{G}(t)$ which fully capture the noise effects. If we impose the  initial conditions $\langle J_x(0) \rangle \equiv \mathcal{J}_x$, $\langle J_y(0) \rangle = 0$, which hold for the CSS and OAT states we are interested in, we are led to the following ratio estimator: 
\begin{equation}
\hat{b}_{\text{R}}^{\text{OAT}} (\tau)= \frac{1}{\tau} \arctan [\langle \hat{J}_y (\tau) \rangle/\langle \hat{J}_x (\tau) \rangle ].
\label{RE2}
\end{equation}
Again, implementing the same number of measurements $\nu= T/\tau $ for each observable doubles the time resources from $T$ to $2T$. Linearizing the expressions in the asymptotic limit $\nu \gg 1$ and high prior knowledge yields the desired expression for the squared uncertainty:
\begin{align}
 \Delta \hat{b}_{\text{R}}^{\text{OAT}}(\tau)^2 & \!=\! 
 \bigg(\!\frac{\partial \hat{b}_{\text{R}}}{\partial \langle J_x(\tau)\rangle}\!\bigg)^2 \!\Delta \hat{J}_x^2(\tau) + \!\bigg(\!\frac{\partial \hat{b}_{\text{R}}}{\partial \langle J_y(\tau)\rangle}\!\bigg)^2 \!\Delta \hat{J}_y^2(\tau)  
 \nonumber \\
& =\frac{\langle J_x (\tau) \rangle^2 \Delta J_y^2(\tau)+ \langle J_y (\tau) \rangle^2 \Delta J_x^2(\tau) }{T \tau [\langle J_x (\tau) \rangle^2 + \langle J_y (\tau) \rangle^2 ]^2 }. 
\label{dbR} 
\end{align}
This provides a ratio estimator useful to carry out asymptotically unbiased quantum sensing in the presence of dephasing, starting from a class of squeezed states with $\langle J_x(0)\rangle \ne 0 $ and $\langle J_y(0)\rangle= 0$. 

\begin{figure*}[t!]
\centering
\includegraphics[width=8cm]{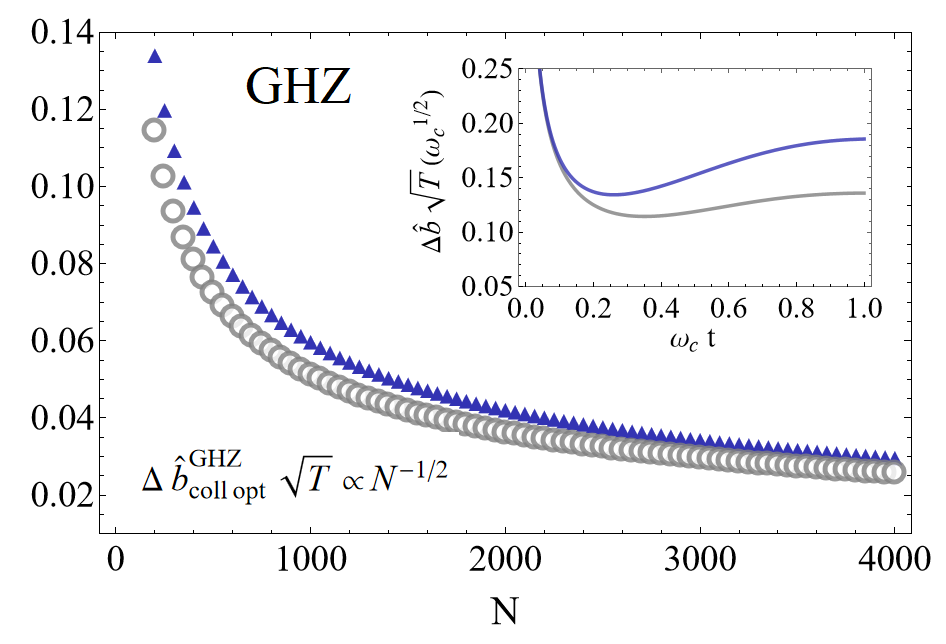}
\hspace*{3mm}
\includegraphics[width=8.4cm]{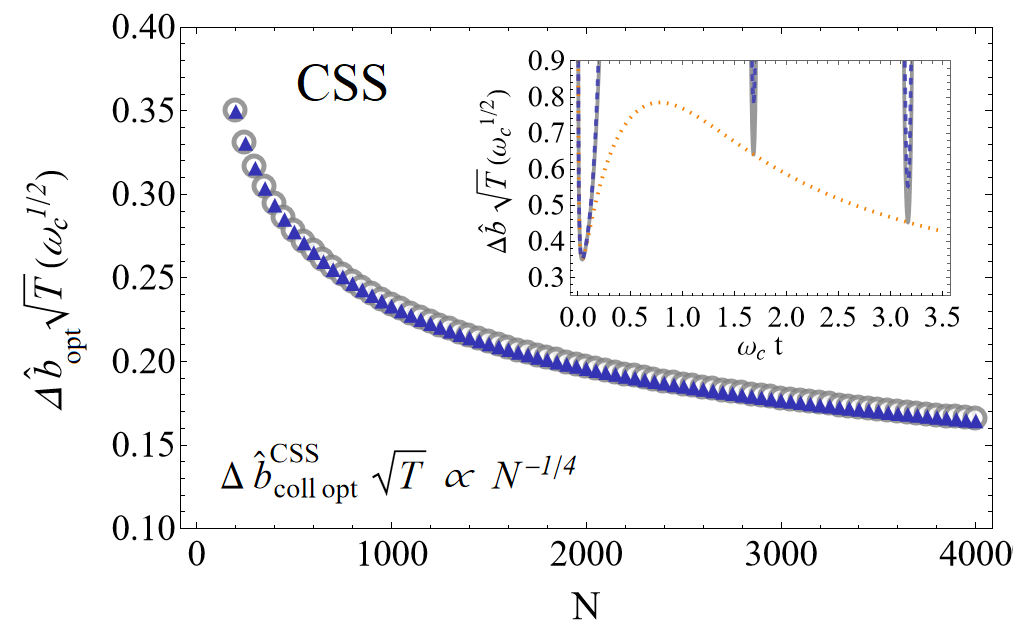}
 \vspace*{-3mm}
\caption{{\bf Ratio estimator comparative performance for collective spin-boson dephasing. } Left: GHZ optimal uncertainty  vs. qubit number for the standard (grey circles) and ratio estimator (blue triangles). Both saturate the SQL, with the standard uncertainty outperforming $\Delta \hat{b}^{\text{GHZ}}_{\text{R opt}}$ by a constant $\text{C}\approx 0.85$ factor. Inset: GHZ $N=100$ qubit uncertainty as a function of time for standard (grey, solid) and ratio estimator (blue, solid). Right: CSS optimal uncertainty vs. qubit number for the standard (grey, circles) and ratio estimator (blue, triangles). Both estimators show $N^{-1/4}$ scaling with the same constant factor when evaluated at their respective optimal phases. Inset: CSS $N=100$ qubit uncertainty as a function of time for standard (grey, solid) and ratio estimator (blue, dashed). The limit of no quantum noise, $\xi(\tau)=0$, for $\Delta \hat{b}(\tau)$ in Eq.\, (\ref{uncs}) (orange, dotted) is included for comparison as a lower bound. {\em Collective} spin-boson dephasing noise from a 1D zero-temperature environment, and a supra-Ohmic spectral density as in Fig.\,\ref{dbplot} is assumed.}
\label{app2} 
\end{figure*}

The expression in Eq.\,(\ref{dbR}) can be simplified by obtaining the $\varphi$ dependence of $\langle J_x^2 (\varphi) \rangle$ and $\langle J_y^2 (\varphi) \rangle$. Taking first- and second-order derivatives with respect to $\varphi$ and solving the resulting partial differential equation system leads to the following mean values: 
\begin{eqnarray*}
   \langle J_x^2(\varphi)\rangle &= \mathcal{K}(t) +\mathcal{A}(t) \cos( 2 \varphi) + \mathcal{B}(t) \sin(2 \varphi),\\
    \langle J_y^2(\varphi)\rangle &= \mathcal{K}(t) - \mathcal{A}(t) \cos( 2 \varphi) - \mathcal{B}(t) \sin(2 \varphi).
\end{eqnarray*}
Here, $\mathcal{K}(t)$, $\mathcal{A}(t)$ and $\mathcal{B}(t)$ are state-dependent quantities that capture the noise effects. Enforcing the initial conditions $\langle J_x^2(0)\rangle = \mathcal{J}_x^2$ , $\langle J_y^2(0)\rangle = \mathcal{J}_y^2$ 
and $ \langle [J_x,J_y]_+ (0) \rangle =0 $, again valid for the CSS and OAT states considered here, we get $\mathcal{B}(t)=0$. 
 Substituting in Eq.\,(\ref{dbR}) yields
\begin{eqnarray*}
\Delta \hat{b}_{\text{R}}^{\text{OAT}}(\tau)^2 \!= \!\frac{2 \mathcal{K}(\tau)- \mathcal{A}[\tau) (1+ \cos(4 \varphi)] -\mathcal{F}(\tau)^2  \sin(2 \varphi)^2 
   }{2 T \tau \mathcal{F}(\tau)^2}.  
\end{eqnarray*}
On the other hand, the precision attained by using the standard method of moments for a set of measurements taking the same fixed total time $2T$ is given by:
\begin{align}
\Delta \hat{b}(\tau)^2 &= \frac{\Delta J_y^2(\tau)}{2T \tau (\partial_{\varphi} \langle J_y(\tau) \rangle)^2} \nonumber \\ 
&=\frac{\mathcal{K}(\tau)+ \mathcal{A}(\tau)\cos(2 \varphi) -(\mathcal{F}(\tau) \sin(\varphi))^2}{2T \tau \mathcal{F}(\tau)^2 \cos(\varphi)^2}. 
\label{dbStd}
\end{align}

Let us now compare the performance of both uncertainties for a product CSS state along $x$ undergoing collective spin-boson dephasing, as considered for the GHZ state. In this scenario the relevant mean values can be computed exactly \cite{FelixPRA,FUR}, from which the quantities $\mathcal{F}(t),\mathcal{K}(t), \mathcal{A}(t)$ are easily deduced. Substituting into Eqs.\,(\ref{dbR})-(\ref{dbStd}) at their respective optimal phases $\varphi=  \pi (2 n +1/4)$ and $\varphi=  n \pi , n \in \mathbb{N}$, for fixed total time $2T$, we find for the CSS squared uncertainties arising from the ratio estimator and the method of moments:
\begin{align}
\Delta \hat{b}_{\text{R}}^{\text{CSS}}(\tau)^2 &=\frac{1}{2 T \tau} \bigg[ \frac{e^{\kappa(\tau)}(N+1) }{2 N\,\cos(\xi(\tau))^{2N-2)} }-1 \bigg], \label{uncs}\\
\Delta \hat{b}^{\text{CSS}}(\tau)^2  &= \frac{(N+1) e^{\kappa(t)}- (N-1) e^{-\kappa(t)} \cos(2 \xi(\tau))^{N-2}}{2 N\, (2T)\,\tau\, \cos(\xi(\tau))^{2N-2}}. 
\notag
\end{align}
For the standard estimation setting, the optimal measurement time and precision for $\Delta \hat{b}^{\text{CSS}}(\tau)$ have been computed to be \cite{FelixPRA}  $ \tau^{\text{CSS}}_{\text{opt, coll}}= (\kappa_0\, \omega_c)^{-1} N^{-1/2}$ and $\Delta \hat{b}^{\text{CSS}}_{\text{opt, coll}} =  (\kappa_0\, \omega_c/T)^{1/2} N^{-1/4} $, respectively. For the ratio estimator, we can carry out an expansion in the short-time limit to derive the optimal measurement time, $\tau^{\text{CSS}}_{\text{R opt, coll}}= (\kappa_0\, \omega_c)^{-1} N^{-1/2}=\tau^{\text{CSS}}_{\text{opt, coll}}$, and performance $\Delta \hat{b}^{\text{CSS}}_{\text{ R opt, coll}}= (\kappa_0\, \omega_c/T)^{1/2} N^{-1/4}$. In line with our findings for the GHZ state, the ratio estimator optimal precision follows the same $N$-scaling as its standard counterpart; furthermore, for a CSS it is multiplied by the {\em same} constant factor and hence attains identical precision: $\Delta \hat{b}^{\text{CSS}}_{\text{ R opt coll}}= \Delta \hat{b}^{\text{CSS}}_{\text{opt coll}}$. This is shown in Fig.\,\ref{app2}, where we plot the optimal uncertainty obtained by numerical optimization as a function of qubit number for both estimators and CSS as well as GHZ initial states evaluated at their respective ideal phases. The results confirm that the analytic predictions for both scaling and constant factors are valid to great accuracy.

\section{Noise-tailored lattice sensor}
\label{sec:lattice}

\subsection{Spin-boson model}
\label{SB}

As a concrete illustrative setting, we consider a 1D dephasing spin-boson model, whereby the qubits interact with a collection of oscillators vibrating at frequencies $\Omega_k$, in thermal equilibrium at inverse temperature $\beta$. Then, 
$$B_n(t)= \sum_k (g_k\, e^{i k r_n} e^{i \Omega_k t}\, b_k^{\dagger} + \text{H.c.}),$$ 
in terms of bosonic operators $b_k$, $b^\dag_k$, with $g_k \in {\mathbb R}$ being a coupling strength for momentum mode $k>0$. We assume a linear dispersion $\Omega_k \equiv v k$, with $v >0$ a speed parameter, and envision the qubits to be arranged in a regular lattice with tunable spacing $x_0$, that is, the position of qubit $n$ obeys $r_n = n x_0$,  $1 \leq n \leq N$. By contrast, the collective limit analyzed in the previous section is obtained when all qubits share the same position, $r_n=x_0,\, \forall n$, leading to \emph{perfect spatial correlations} which cannot be controlled. For a qubit pair $n,m$, a \emph{transit time} proportional to their distance may then be defined by $t_{nm} \equiv |n-m| x_0 /v $. 

\begin{figure*}[t!]
\centering
\includegraphics[width=8cm]{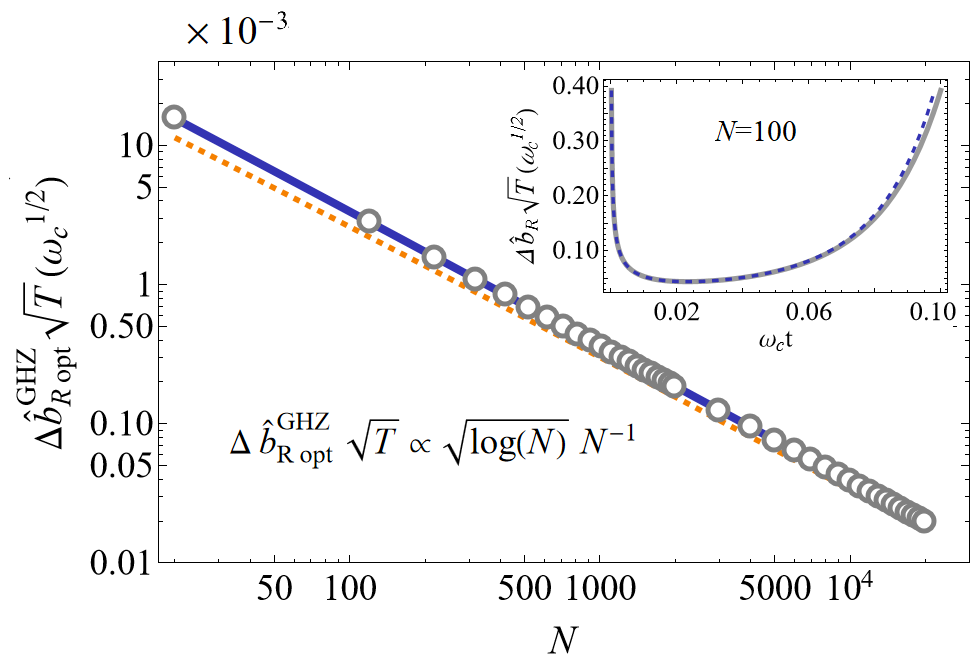}\hspace*{3mm}\includegraphics[width=8.4cm]{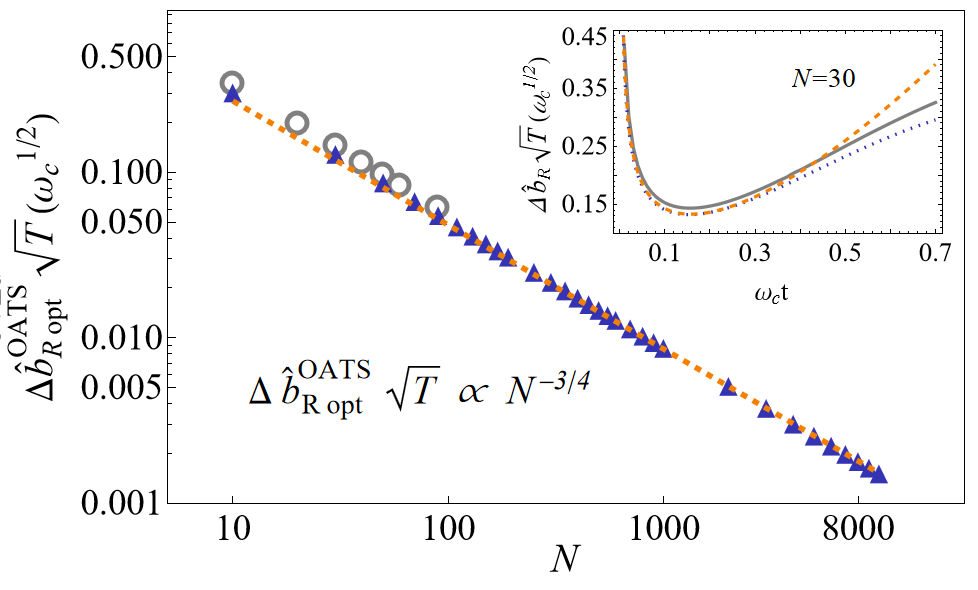}
\vspace*{-3mm}
\caption{{\bf Noise-optimized superclassical precision scaling.} Left: GHZ optimal uncertainty vs. qubit number. Grey circles: Exact numerical optimization. Blue, solid line: Analytic expression for $\Delta \hat{b}^{\text{GHZ}}_{\text{opt}}$. Orange, dashed: Asymptotic scaling limit, Eq.\,\eqref{GHZsc}. Inset: GHZ uncertainty vs. time for $N=100$ qubits and optimal lattice spacing $x_0=0.4296$. Grey, solid: Exact expression. Blue, dashed: Short-time approximate uncertainty. Right: Optimal uncertainty vs. qubit number for an OATS  ideally squeezed along $y$ before evolution. Blue triangles: Numerical optimization for purely classical noise. Gray circles: Numerical optimization including quantum noise. Orange, dashed: Asymptotic scaling limit, Eq.\,\eqref{eq:oat}. Inset: OATS uncertainty vs. time for $N=30$ qubits and optimal spacing $x_0=0.46$. Blue, dashed: Purely classical dephasing. Grey, solid: Classical and quantum dephasing. Orange, dotted: Short-time approximate uncertainty. All other parameters are as in Fig.\,\ref{dbplot}. }
\label{both}
\end{figure*}

Notably, for the above setting we have $\varphi_1(t)\equiv 0$ under standard physical assumptions on the coupling \cite{IschiPRB,FUR}. The dynamic coefficients determining $\gamma(t)$ and $\varphi_0(t)$ are given in Eq.\,\eqref{eq:decay} and Eq.\,\eqref{phi0}, respectively. To compute $\kappa_{nm}(t)$ and $\xi_{nm}(t)$, we further assume a continuum of bosonic modes, characterized by a spectral density of the form 
\begin{equation}
J(\omega) \equiv \alpha \omega_c (\omega/\omega_c)^s e^{-\omega/\omega_c}, \quad \alpha >0,\: s>0,
\label{sd}
\end{equation}
with $s$ being the so-called Ohmicity parameter. In a low-temperature regime where thermal effects may be taken to be negligible ($\coth(\beta|\omega|/2\approx 1$), the noise classical and quantum spectra are then calculated as 
$$S^+_{nm}(\omega) = 4 \pi J(\omega) \cos(\omega t_{nm}), \;\; S^-_{nm}(\omega) = S^+_{nm}(\omega) \text{sgn}(\omega),$$
where $\text{sign}(\omega)$ is the sign function. 

The cutoff frequency $\omega_c$ defines a metrologically relevant short-time limit, given by $\omega_c t \ll 1$, where the quantum advantage in estimation precision may be maximized \cite{Chin2012,FelixPRA,FUR}. In this regime, the dynamic coefficients entering $\gamma(t)$ and $\varphi_0(t)$ can be approximated as 
\begin{eqnarray*}
\kappa_{nm}(t) & \approx & \kappa_0^2 (\omega_c t)^2 \delta_1(|n-m| x_0), \quad \kappa_0^2\equiv\alpha \,\Gamma({s}+1),\\
\xi_{nm}(t)  & \approx &\xi_0^3 (\omega_c t)^3 \delta_2(|n-m|x_0), \quad \xi_0^3\equiv({\alpha}/{6}) \, \Gamma (s+2),
\end{eqnarray*}
with $\Gamma$ being the Euler Gamma function. Their temporal dependence is thus factored from the relevant spatial correlations, which are encoded in the functions 
\begin{eqnarray}
\delta_1(|n-m| x_0)& \equiv &\;u^{{s}+1}\, T_{{s}+1}(u), 
\label{spatial1} \\
\delta_2(|n-m| x_0)& \equiv & \;u^{s+2}\, T_{s+2}(u),
\label{spatial2}
\end{eqnarray} 
where $u\equiv[1+( |n-m| x_0)^2]^{-1/2}$ and one can show that $T_{n}(x)$ denotes the $n$-th order Chebyshev polynomial of the first kind. Provided that knowledge of the spectral density $J(\omega)$, hence of the spatial correlation functional forms $\delta_\ell (x)$ is available, and that $J(\omega)$ exhibits supra-Ohmic low-frequency behavior ($s>1$), spatial correlations with $n\ne m$ can be made \emph{negative} by tuning the distance $x_0$. We now show how this may engender a drastic improvement in the scaling of estimation precision with respect to both collective noise \cite{FelixPRA} and our previous randomized Ramsey protocol \cite{FUR}.

\subsection{Initial GHZ state}

\begin{figure*}[t!]
\centering
\includegraphics[width=8cm]{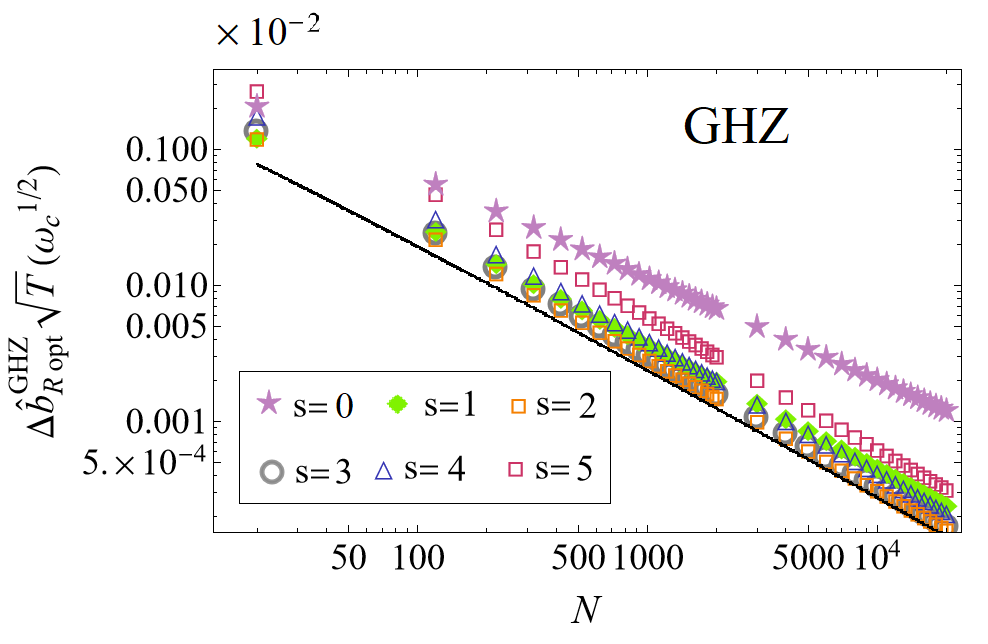}
\hspace*{3mm}
\includegraphics[width=7.8cm]{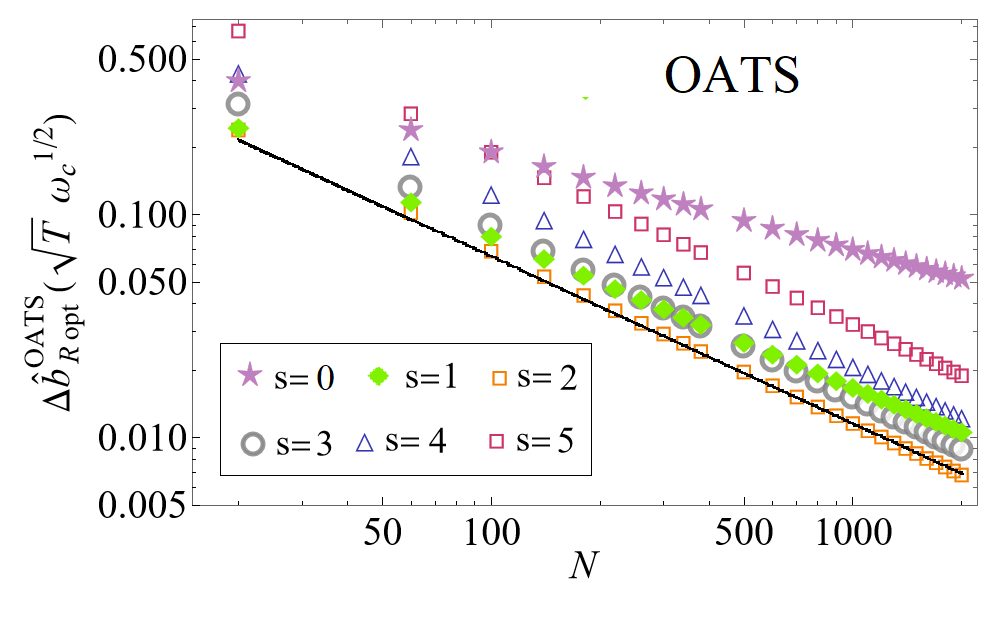}
\vspace*{-3mm}
\caption{{\bf Ratio estimator uncertainty optimal performance for different Ohmicity parameters. } Minimal uncertainty as function of $N$ of a GHZ (left) and OAT state (right) for  different spectral density with a high-frequency exponential cutoff and different Ohmicity parameters $s=0$, (solid lilac stars) $s=1$ (solid green diamonds), $s=2$ (hollow orange squares), $s=3$ (hollow grey circles), $s=4$ (hollow blue triangles), and $s=5$ (hollow brown rectangles). The solid black line describes the asymptotic lower bound of $3^{1/2} \Gamma(3)^{1/4} (\omega_c/T)^{1/2} N^{1/2}$ and $ 3/4 N^{-1} \sqrt{\log(N)} $ for OATS and GHZ, respectively. Here, $\alpha=1=\omega_c$, as in previous figures.
    }
\label{diffs}
\end{figure*}

Since, despite being non-collective, the noise remains pure dephasing, the same analysis carried out in Sec.\,\ref{RGHZ} holds here, thus we may estimate the target frequency without incurring in noise-induced bias through Eq.\,(\ref{RE}). The resulting precision, however, is different, with the dephasing rate $\gamma_{\text{GHZ}}(t)$ no longer being that of the collective setting. To determine the GHZ optimal precision, we not only minimize the uncertainty with respect to time, but also the lattice parameter $x_0$, for optimal phase $N b t=( 2k+1)\, \pi/4$, $k \in \mathbb{N}$. Since $\varphi_0(t) \equiv 0$ for $|\text{GHZ}\rangle$ by symmetry, quantum correlations do not play any role in this case ($\delta_2\equiv 0$ in Eq.\,\eqref{spatial2}). Taking advantage of the fact that, in the short-time limit, 
$$\gamma_{\text{GHZ}}(t)\!\approx\! (\omega_c t)^2 F_N(x_0), \;\:F_N(x_0) \equiv  \kappa_0^2  \sum_{n,m}^N \delta_1(x_0|n-m|),$$ 
the minimization may be carried out with respect to each variable separately. Substituting in Eq.\,(\ref{dbGHZ}) and optimizing with respect to $\tau$ leads to the optimal measurement time, $\tau_{\text{R opt}}^{\text{GHZ}}= \frac{1}{2} \omega_c^{-1}  F_N(x_0)^{-1/2}$. It follows that the best sensing performance is achieved by minimizing the spatial function $F_N(x_0)$ for the time-optimized uncertainty, leading to 
$$\Delta \hat{b}_{\text{R}}(\tau^{\text{GHZ}}_{\text{R opt}}) \approx 2.96\; {\omega_c/T}^{1/2}\, F_N(x_0)^{1/4}\, N^{-1}.$$ 
While the details are provided in Appendix\,\ref{GHZstate} (see in particular Eq.\,\eqref{topt} for the origin of the above numerical prefactor), the resulting approximate (analytically) minimized spatial function, $F_N(x_{0 \text{ opt}}^{\text{GHZ}})$, can be shown to scale logarithmically in the large-$N$ limit: $F_N(x_{0 \text{ opt}}^{\text{GHZ}}) \propto \mathcal{O}(\log(N)^2)$. Accordingly, the optimal asymptotic sensing performance is 
\begin{equation}
\Delta \hat{b}^{\text{GHZ}}_{\text{R opt }} \approx 2.96\, (\omega_c/T)^{1/2}\sqrt{\log(N)}\, N^{-1},  
\label{GHZsc}
\end{equation}
which is \emph{closer to Heisenberg scaling than any power law}. 

Figure \ref{both}(left) demonstrates that the agreement between the analytic expression in Eq.\,\eqref{GHZsc} and the exact numerical optimization over $x_0, \tau$ is excellent even at finite $N \gtrsim 20$, despite the fact that discrepancies between the lattice parameter that numerically minimizes $\delta_1$ and its analytic approximation $x_{0 \text{ an}}^{\text{GHZ}}$ vanish only asymptotically in $N$. This indicates a high degree of robustness against deviations from the exact optimal lattice parameter and, in turn, against uncertainty in the characterization of the underlying noise spectral density.

\subsection{Initial OAT state}

Unlike GHZ states, OATS are not immune to the genuinely quantum contribution of the noise. For clarity, however, let us first consider the classical decay contribution, described as before by Eqs.\,\eqref{eq:decay}-\eqref{freqchi}, and assess the impact of the phase $\varphi_0(t)$ at a later stage. As detailed in Appendix\,\ref{optOAT}, by performing a cumulant expansion over appropriate qubit operators \cite{FelixPRA}, we may derive approximate expressions for the moments $\langle J_v(t)\rangle$, $\langle J_v^2(t)\rangle$, $v=x,y$, which are remarkably accurate in the short-time limit $\omega_c t \ll 1$. Substituting in Eq.\,(\ref{dbR}), it is then possible to expand the uncertainty in powers of $t$, 
$$\Delta\hat{b}_{\text{R}} (t) \approx \big(\sqrt{T t}\, h_0(N) \big)^{-1} \!\bigg(\sum_{k=0,1,2} \! a_{2k}(N, x_0)\, (\omega_c t)^{2k}\bigg)^{\!1/2},$$ 
where the coefficients $h_0(N)$ and $a_{2k}(N,x_0)$ can be evaluated analytically by the same approach used to compute the spatial function $F_N(x_0)$ for the GHZ state. This expansion is accurate in the region where the minimum occurs (see inset of Fig.\,\ref{both}(right)), and once again it allows one to separate the spatial and temporal dependence of the uncertainty into two distinct contributions. The best performance can then be extracted by minimizing the expansion with respect to time and, subsequently, with respect to the lattice parameter. For $s>1$, as detailed in Appendix\,\ref{optOAT}, we find that the noise-tailored sensor reaches the Zeno limit, 
\begin{equation}
\Delta \hat{b}_{\text{R opt}}^{\text{OATS}} \approx {{\omega_c}/{(3T)}}^{1/2}  \,\Gamma(s+1)^{1/4} N^{-3/4}.
\label{eq:oat}
\end{equation}
This is a mere $N^{-1/12}$ scaling loss with respect to the noiseless OAT scenario, and is found to be in full agreement with numerical minimization, see Fig.\,\ref{both}(right). 

Accounting for the effect of quantum noise due to $\varphi_0(t)$ makes the expression for the mean values significantly more involved. This prevents us from deriving a tractable short-time expansion for the uncertainty and lengthens the computational time required to evaluate $\Delta \hat{b}_{\text{R}}^{\text{OAT}}(\tau)$ numerically, limiting in turn the accessible values of $N$. Still, as one may see from the inset in Fig.\,\ref{both}(right), the uncertainty as a function of time has a nearly identical behavior as in the presence of classical noise alone, with the contribution of $\varphi_0(t)$ entering as a small correction. In line with the above, numerical minimization in the range of $N$ we were able to test leads to similar values for the optimal uncertainty as when quantum noise was disregarded, see Fig.\,\ref{both}(right). Such a behavior is plausible in light of the fact that, as in the GHZ case, $\xi_{nm}(t) \propto (\omega_c t)^3$, as compared to $\kappa_{nm}(t) \propto (\omega_c t)^2$, which causes the decay contributions to dominate in the short-time limit. Altogether, this strongly suggests that a Zeno limit as in Eq.\,(\ref{eq:oat}) is realized asymptotically, also when quantum noise is fully included.

\subsection{Parameter robustness}

While the calculations reported thus far for the spin-boson setting have been carried out by assuming a spectral density with Ohmicity parameter $s=3$ and an exponential high-frequency decay, that is, $J(\omega)= \alpha\, \omega_c^{-2}\, \,\omega^3\, e^{-\omega/\omega_c}$, it is important to assess how sensitively the resulting precision scaling depend upon the details of the noise spectra. Numerical optimization indicates that the same nearly-Heisenberg uncertainty scaling and Zeno limit derived for GHZ and OAT states, respectively, hold for other supra-ohmic densities, with the $N$-independent prefactor increasing with $s$, see Fig.\,\ref{diffs}. Furthermore, a linear fit shows that for OAT states, the prefactor is $\Gamma(s+1)/\sqrt{3}$, which was only rigorously derived for $s=3$.  

Figure\,\ref{diffs} also shows that for an Ohmic ($s=1$) spectral density, however, the optimal ratio uncertainty experiences a small scaling loss for both GHZ (left) and OAT (right) states, which manifests as a slight slope decrease with respect to the supra-Ohmic cases in the log-log plot. Finally, for a sub-Ohmic spectral density ($s=0$), such a change in slope and scaling loss are more pronounced, especially for OAT states. This is in line with the expectation that negative spatial correlations play a diminished role in the presence  of non-vanishing zero-frequency noise. Understanding in more detail the precision achievable in the Ohmic and sub-Ohmic scenarios, as well as addressing spectral densities with non-exponential (e.g., Gaussian) high-frequency cutoffs or non-trivial spectral features (e.g., high-frequency peaks), remains a task worth of additional investigation.

\section{Conclusion}
\label{conclusion}

We have shown how to construct ratio estimators for entanglement-assisted frequency estimation by Ramsey interferometry which, at the cost of doubling the required measurement resources, overcome noise-induce bias and retain the same precision scaling of standard local estimators in the asymptotic large-$N$ regime. While our construction is applicable to both Markovian and non-Markovian dephasing settings, and requires no prior detailed noise knowledge, we have additionally shown that asymptotically unbiased frequency estimation with noise-optimized superclassical precision is possible by suitable sensor's engineering and tuning of the noise spatial correlations. In particular, by focusing on spin-boson dephasing in a one-dimensional lattice geometry with tunable qubit-probe separation, we established a novel scaling which is a $\log(N)^{1/2}$ factor away from the Heisenberg limit for a GHZ state, and found that Zeno scaling is reachable with a properly squeezed OAT state.

An outstanding question for future work is to explore potential experimental realizations of our tunable lattice-sensor model, for instance in trapped ion settings \cite{monroe,postler} or NV centers in diamond \cite{Zhou}. From a theoretical standpoint, our results hint at the possibility that the Heisenberg limit may be reachable by further optimizing the initial entangled state, in addition to the sensor's geometry. Likewise, it would be interesting to extend the applicability of our noise-unbiased estimators to generalized Ramsey echo-based protocols,  for which the symmetry condition $\langle J_y(0)\rangle =0$ of the input states is relaxed, and pre-measurement anti-squeezing operations are allowed \cite{SchulteEchoes,Hammerer}. Finally, assessing the usefulness of our ideas in a non-asymptotic metrology setting with a limited amount of data \cite{Rubio, Meyer}, where a Bayesian estimation paradigm is appropriate, and noise-induced bias competes with finite-sample effects, may be especially important to expand the practical reach of quantum metrology applications.

\vfill

\section*{Acknowledgements}

It is a pleasure to thank Augusto Smerzi for valuable input and clarifications, and F\'elix Beaudoin for a critical reading of the manuscript. L.V. also acknowledges early discussions with F\'elix Beaudoin on the issue of noise-induced bias. Work at Dartmouth was supported by the US NSF through Grant No.\,PHY-2013974, and the Constance and Walter Burke Special Projects Fund in QIS. Work at Griffith was supported (partially) by the Australian Government through the Australian Research Council's Discovery Projects funding scheme (project No. DP210102291).

\appendix* 

\onecolumngrid

\section{Additional detail on technical derivations}
\label{app:lattice}

\subsection{Optimal noise-tailored precision scaling for initial GHZ states} 
\label{GHZstate}

Here we derive the GHZ optimal uncertainty achievable by our ratio estimator in the presence of spin-boson noise with tunable spatial correlations. Our starting point is Eq.\,\eqref{dbGHZ}, evaluated for optimal phase $\varphi= \pi/4$ in the short-time regime:
\begin{equation}
\Delta \hat{b}_{\text{R}}^{\text{GHZ}}(\tau)^2 
= \frac{1}{N^2 T \tau}\bigg( e^{2 \gamma_{\text{GHZ}}(\tau) } -\frac{1}{2} \bigg)
\approx \frac{1}{N^2 T \tau} \bigg( e^{2 \kappa_0^2 (\omega_c \tau)^2 \sum_{n,m=1}^N \delta_1(x_0|n-m|)  } - \frac{1}{2} \bigg).
\label{dbGHZST}
\end{equation}
One may first minimize with respect to time, leading to the optimal measurement time and the corresponding time-optimized uncertainty. The result may be cast in the following form: 
\begin{equation}
\tau_{\text{R opt}}^{\text{GHZ}}= \frac{1}{2} \omega_c^{-1}  F_N(x_0)^{-1/2} \sqrt{1+\frac{1}{2} W(e^{-1/2}/4) } ,\quad \:\Delta \hat{b}_{\text{R}}(\tau_{\text{R opt}}^{\text{GHZ}}) = \sqrt{2}\,\frac{(1+2\, W(e^{-1/2}/4) )}{\sqrt{-W(e^{-1/2}/4)}}\;\sqrt{\frac{\omega_c}{T}}\; F_N(x_0)^{1/4}\; N^{-1},
\label{topt}
\end{equation}
where $F_N(x_0)=\kappa_0^2\, \sum_{n,m=1}^N \delta_1(x_0 |n-m|)$ is the function of the lattice spacing $x_0$ also defined in the main text, and $W(x)$ is the principal branch of the Lambert's $W$ function \cite{Corless}. Recall that the Lambert's $W$ function $W_k(z)$ is defined, given complex numbers $w$ and $z$, by the equation $w e^w=z$, which holds if and only if $w=W_k(z)$ for some integer $k$. For simplicity, we have dropped the subscript $k$ for the principal branch $k=0$. The numerical prefactor in Eq.\,(\ref{topt}) can be computed to be $\sqrt{2}\;(1+2\, W(e^{-1/2}/4)) (-W(e^{-1/2}/4))^{-1/2} \approx 2.96$, as quoted in the main text above Eq.\,\eqref{GHZsc}. 

In order to obtain the best precision, we must now minimize $F_N(x_0)$ with respect to $x_0$. Using the symmetry of a 1D regular lattice, we may write the spatial function $F_N(x_0)$ in terms of a single index sum, and a term including the spatial ``self-correlations'' $\delta_1(0)$ corresponding to the $n=m$ case, that is, 
\begin{equation}
 F_N(x_0)=\kappa_0^2 \bigg[ N \delta_1(0)+ 2 \sum_{j=1}^{N-1} (N-j) \delta_1(j\, x_0)\bigg]. 
 \label{FN} 
\end{equation}
Replacing the explicit functional form of the spatial correlations $\delta_1(|n-m| x_0)$ for $s=3$ into $F_N(x_0)$ yields
\begin{equation}
F_N(x_0)= \kappa_0^2 \bigg[ N + 2 \sum_{j=1}^{N-1} (N-j)\, \frac{1- 6\, (j\,x_0)^2+ (j\,x_0)^4}{(1+(j\,x_0)^2)^4} \bigg]=\kappa_0^2 \left[ N +S_N(x_0)\right] ,
\label{FNb}
\end{equation}
with $\kappa_0^2=\alpha\, \Gamma(4)$ a dimensionless constant and $S_N(x_0)$ consisting of two finite sums:
$$ S_N(x_0)= 2 N \sum_{j=1}^{N-1} \, \frac{1- 6\, (j\,x_0)^2+ (j\,x_0)^4}{(1+(j\,x_0)^2)^4} -2 \sum_{j=1}^{N-1} j\, \frac{1- 6\, (j\,x_0)^2+ (j\,x_0)^4}{(1+(j\,x_0)^2)^4}. $$

To evaluate these series, we now show that they can be written as linear combinations of Polygamma functions, defined by $\psi^{(m)}(z) \equiv (-1)^{m+1} \int_0^{\infty} \frac{t^m e^{-z t}}{1-e^{-t}} \, dt $. The key step is to expand $(1-e^{-t})^{-1}$ in the integrand that defines $\psi^{(m)}(t)$ by using the geometric series, and summing each contribution term-by-term. For example, for 
$$\text{Re}\,\psi^{(3)}\left( \frac{i+ x_0}{x_0}\right)=\frac{1}{2}\left[\psi^{(3)}\!\left( \frac{i+ x_0}{x_0}\right)+\psi^{(3)}\!\left( \frac{-i+ x_0}{x_0}\right)\right],$$ we have 
\begin{eqnarray}
\text{Re}\,\psi^{(3)}\bigg( \frac{i+ x_0}{x_0}\bigg) \!= \! \int_0^{\infty} \!\frac{t^3 e^{-t}}{1-e^{-t}} \cos\Big(\frac{t}{x_0}\Big)  dt = \sum_{j=1}^{\infty} \int_0^{\infty} \!\!t^3 \; e^{-j t} \cos\Big(\frac{t}{x_0}\Big) dt 
 = 6 x_0^4\; \sum_{j=1}^{\infty} \frac{1- 6\, (j\,x_0)^2+ (j\,x_0)^4}{(1+(j\,x_0)^2)^4}, \qquad
 \label{Re1}
\end{eqnarray}
where integrating each series coefficient independently is allowed by Fubini's theorem. Similarly, 
\begin{eqnarray}
\text{Re}\;\psi^{(3)}\bigg( \frac{i}{x_0}+N\bigg) = 6 x_0^4  \sum_{j=N+1}^{\infty} \frac{1- 6\, (j\,x_0)^2+ (j\,x_0)^4}{(1+(j\,x_0)^2)^4}. \label{Re2}
\end{eqnarray} 
We can then subtract Eq.\,(\ref{Re2}) from Eq.\,(\ref{Re1}) and, multiplying by $ N\, \kappa_0^2 \,x_0^{-4}$, we get:
\begin{eqnarray}
N\, \kappa_0^2 \,x_0^{-4} \left[\text{Re}\;\psi^{(3)}\left( \frac{i+ x_0}{x_0}\right)- \text{Re}\;\psi^{(3)}\left( \frac{i}{x_0}+N\right) \right]  =2 N \sum_{j=1}^{N-1} \, \frac{1- 6\, (j\,x_0)^2+ (j\,x_0)^4}{(1+(j\,x_0)^2)^4},
\end{eqnarray}
which is the first of the desired sums in $S_N(x_0)$.  The second sum can be computed in a similar fashion by combining the appropriate Polygamma functions. We obtain, carrying out the algebra:
\begin{eqnarray}
S_N(x_0)&= &\frac{N}{6 x_0^5} \bigg \{ 3 x_0 \, \text{Re}\;  \psi^{(2)}\left( \frac{i+ x_0}{x_0}\right) - 3 x_0\, \text{Re}\;  \psi^{(2)}\left( \frac{i}{x_0}+N\right) - 2\, \text{Im}\;  \psi^{(3)}\left( \frac{i+ x_0}{x_0}\right) \nonumber\\
&+ & x_0\, N \,\text{Re}\;  \psi^{(2)}\left( \frac{i+ x_0}{x_0}\right) + \, \text{Im}\; \psi^{(3)}\left( \frac{i}{x_0}+N\right) +\, x_0 N\, \text{Re}\; \psi^{(3)}\left( \frac{i}{x_0}+N\right) \bigg \}.
\label{SN} 
\end{eqnarray}
Eq.\,(\ref{SN}) can be further simplified by invoking two key properties of the Polygamma functions:
\begin{eqnarray*}
\psi^{(m)}(z+1)= \psi^{(m)}(z)- (-1)^m m!\; z^{-m-1}, \qquad 
\psi^{(m)}(z)-(-1)^m \psi^{(m)}(1-z) = - \pi \frac{d^m}{d z^m} \cot(\pi z).
\end{eqnarray*}
This allows us to write some of the terms in Eq.\,(\ref{SN}) in terms of elementary functions, by using
\begin{eqnarray*}
\psi^{(m)}\!\left( \frac{i+ x_0}{x_0}\right) - (-1)^{m} \psi^{(m)}\!\left( \frac{-i+ x_0}{x_0}\right)= (-1)^m m! \left(\frac{i}{x_0} \right)^{-m-1} \hspace*{-4mm}- \pi \frac{d^m}{d u^m} \cot(\pi u),
\end{eqnarray*}
with $u=x_0^{-1}$. For example, recalling the above expression for $\text{Re}\; \psi^{(3)}\left( \frac{i+x_0}{x_0}\right)$, we have  
\begin{eqnarray*}
\text{Re}\; \psi^{(3)}\!\left( \frac{i+x_0}{x_0}\right) = -3 + \left(\frac{\pi}{x_0}\right)^4 \left[2+ \cosh \left(2 \frac{\pi}{x_0}\right) \right] \sinh \left( \frac{\pi}{x_0} \right)^{-4} \approx -3 +8 \left(\frac{\pi}{x_0}\right)^4 e^{  -2 {\pi}/{x_0} }, 
\end{eqnarray*}
where the last approximate equality comes from assuming a small-lattice parameter regime, $x_0 \lesssim 0.5$, which is where the optimal qubit separation lies, as shown by numerical optimization. Furthermore, all terms carrying Polygamma functions with $({i}/{x_0}+N)$ in their argument give a negligible contribution to the series $S_N(x_0)$ in the region where the minimum of $F_N(x_0)$ occurs. We can then safely drop them in Eq.\,(\ref{SN}), if our task is to find the optimal lattice separation. Lastly, we expand the remaining terms in  Eq.\,(\ref{SN}) in powers of $x_0$, for $x_0 \lesssim 0.5$. Substituting  in Eq.\,(\ref{FN}) after performing these steps finally leads to the following analytic approximation to the spatial function:
\begin{eqnarray}
F_N(x_0) \approx \frac{1}{6} \bigg[ 8 N \left(\frac{\pi}{x_0}\right)^4 \exp \left(-2 \frac{\pi}{x_0} \right)+ 2\, \frac{1}{\pi^2}\, \left(\frac{\pi}{x_0}\right)^2 \bigg]\equiv F^{\text{an}}_{N}(x_0).
\label{FNan}
\end{eqnarray} 

\begin{figure}[t!]
    \centering
    \includegraphics[width=15cm]{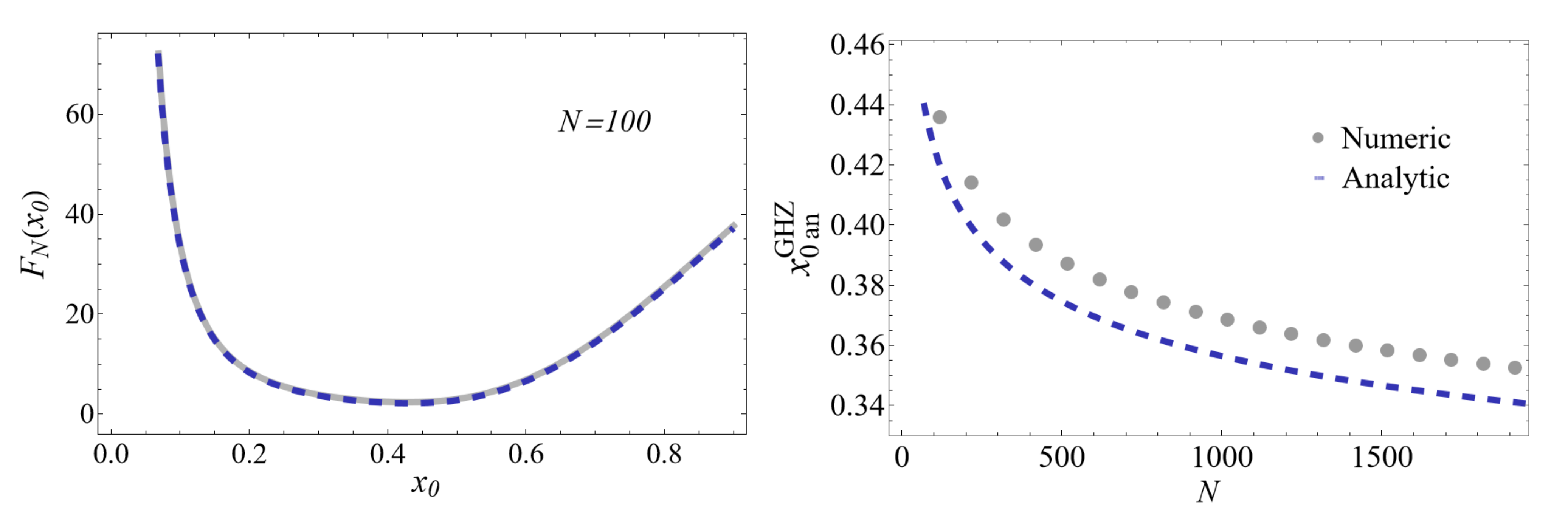}
    \vspace*{-3mm}
    \caption{{\bf Spatial function and optimal lattice parameter.} Left: Spatial function $F_N(x_0)$ given by Eq.\,(\ref{FN}) (grey, solid) and analytic approximation $F^{\text{an}}_N(x_0)$ given by Eq.\,(\ref{FNan}) (blue, dashed) vs. lattice parameter $x_0$ for $N=100$ qubits. Right: Optimal lattice parameter as a function of qubit number. Grey dots: Exact numerical optimization, yielding the optimal lattice parameter $x^{\text{GHZ}}_{0\text{ opt}}$. Blue, dashed: Approximate analytic solution $x^{\text{GHZ}}_{0 \text{an}}$, given by Eq.\,(\ref{xopt}). Noise parameters as in Fig.\,\ref{dbplot}, except that now dephasing is clearly not collective.}
    \label{sp}
\end{figure}

In Fig.\,\ref{sp}(left) we have plotted $F_N(x_0)$ and $F^{\text{an}}_{N}(x_0)$ as a function of the lattice parameter $x_0$ for $N=100$ qubits. One can see that the analytical expression $F^{\text{an}}_{N}(x_0)$ reproduces the behavior of the spatial function to great accuracy, which only increases for large $N$. Bearing that in mind, in what follows we drop the superscript ``an'' and denote it by the same label as its exact counterpart, $F_N(x_0)$. The optimal lattice parameter minimizing $F_N(x_0)$ lies around $x_{0 \text{ opt}}^{\text{ GHZ}}\approx 0.44$; however, there is a neighborhood around this point in which the spatial function stays approximately constant. Importantly, since small departures from the optimal $x_0$ value do not drastically alter $F_N(x_0)$, no prohibitively precise noise characterization is needed. This grants our noise-tailored sensor a degree of parameter robustness with respect to deviations from the ideal lattice separation. 

To obtain an approximate expression for $x_{0 \text{ opt}}^{\text{GHZ}}$, it is convenient to let $v\equiv {\pi}/{x_0}$ and write $F_N(v)=\frac{1}{6} \left( 8 N v^4 e^{ -2 v}+ 2 \frac{1}{\pi^2}\, v^2 \right)$. Taking the derivative with respect to $v$ and equating to zero then yields the following condition:
\begin{eqnarray}
8 N e^{-2 v} v^2 (2-v) + \frac{1}{\pi^2} =0 \quad \Rightarrow \quad
 -8 \pi^2 N  e^{-2 v} v^3 \approx 1  . 
\label{v}
\end{eqnarray}
The second approximate equality is fulfilled in the $N \gg 1$ limit, where the optimal lattice  parameter obeys $x_{0 \text{ opt}}^{\text{ GHZ}} \ll 1$, and hence $v^{\text{GHZ}}_{\text{ an}} \gg 2$. This can again be solved with the help of the Lambert $W$ function. The relevant solution is given by
\begin{eqnarray}
 v^{\text{GHZ}}_{\text{ an}}= \frac{\pi}{x_{0 \text{ opt}}^{\text{GHZ}}} \approx -\frac{3}{2}\; W_{-1}\!\left( -(3\; \pi^{2/3}\; N^{1/3})^{-1} \right).
 \label{vopt}
\end{eqnarray}
Since the argument in the Lambert function vanishes for $N \gg 1 $, we can use an expansion for $W_{-1}(x)$, valid when $x<0$ and $|x| \ll 1$ \cite{Corless}, to cast Eq.\,(\ref{vopt}) into a simpler form. That is, $W_{-1}(x) \approx L1 - L2 + \frac{L1}{L2}$, with $L1\equiv \log(|x|)$, and $L2\equiv \log(-\log(|x|))$. In practice, keeping the first two terms of the expansion, $W_{-1}(x) \approx L1-L2$, greatly simplifies the ensuing algebra and still provides an accurate enough characterization for our purposes. Substituting in Eq.\,(\ref{vopt}) and solving for $x_{0 \text{ an}}^{\text{GHZ}}$, we finally obtain 
\begin{eqnarray}
x_{0\text{ opt}}^{\text{GHZ}} \approx x_{0\text{ an}}^{\text{GHZ}} =  \frac{2 \pi}{ \log(3\pi^{2/3} N^{1/3} )+ \log(\log(3\pi^{2/3} N^{1/3} ))}.
\label{xopt}
\end{eqnarray} 

We have plotted the analytic approximation $x_{0\text{ an}}^{\text{GHZ}} $ and exact optimal lattice parameter $x_{0\text{ opt}}^{\text{GHZ}}$, obtained by numerical optimization, as function of qubit number $N$ in Fig.\,\ref{sp}(right). We can see that Eq.\,(\ref{xopt}) has an $N$-dependent offset. This is the result of having discarded the constant factor, $(v-2) \approx v$ in Eq.\,(\ref{v}) when minimizing the spatial function, which only vanishes in the asymptotic limit. Despite this, the analytic value lies close enough to the true value of the optimal lattice parameter to yield an excellent approximation to the optimal uncertainty, see Fig.\,\ref{both} in main text. Upon substituting Eq.\,(\ref{xopt}) into Eq.\,(\ref{FNan}), we arrive at an expression for the minimum of the spatial function, namely, 
\begin{eqnarray}
F_N(x_{0 \text{ an}}^{\text{ GHZ}}) = \frac{1}{{6 \pi ^2}} \Big[\log \left( \pi ^2 N\right)+\log \left(\log ^3\left(27 \pi ^2 N\right)\right)\Big]^2 \bigg[\frac{\left(\log \left(\log ^3\left(27 \pi ^2 N\right)\right)+\log \left( \pi ^2 N\right)\right)^2}{\log \left(27 \pi ^2 N\right)}+\frac{1}{2}\bigg]
\label{FNanopt},
\end{eqnarray}
which is of $\mathcal{O}(\log(N)^2)$ in the $N\gg 1$ limit. Finally, replacing $F_N(x^{\text{GHZ}}_{0 \text{ an}})$ in the time-optimized uncertainty, Eq.\,(\ref{topt}), we obtain an analytic formula for the overall minimun uncertainty, 
\begin{eqnarray}
\Delta \hat{b}^{\text{GHZ}}_{\text{R opt}} (N) \approx   \sqrt{2}\;\frac{(1+2\, W(e^{-1/2}/4) )^{1/4}}{\sqrt{-W(e^{-1/2}/4)}}\;\sqrt{\frac{\omega_c}{T}}\; \left[ F_N(x^{\text{GHZ}}_{0 \text{ an}})\right]^{1/4} N^{-1} \propto \sqrt{\log(N)}\,N^{-1} , 
\end{eqnarray}
with  $\sqrt{2}\;(1+2\, W(e^{-1/2}/4) )^{1/4} (\sqrt{-W(e^{-1/2}/4)})^{-1} \approx 2.96 $  as noted above, and with the last proportionality relation holding for $N\gg 1$. Remarkably, the agreement between the above analytic expression and the exact numerical result is excellent even at finite number of qubits, $N \gtrsim 20$, despite the discrepancies between $x^{\text{GHZ}}_{0 \text{ opt}}$ and  $x_{0\text{  an}}^{\text{GHZ}}$; see again Fig.\,\ref{both} in the main text.

\subsection{Initial OAT states}
\label{OATs}

\subsubsection{Expectation values of collective angular momentum operators} 
\label{expect}

Generally, the dynamics of a state $|\text{OATS}\rangle= e^{-i \beta J_x} e^{-i \theta J_z^2}|\text{CSS}\rangle_x$ subject to Gaussian dephasing noise are highly entangled, and exact expressions for the relevant mean values are not available. To evaluate the mean values, a cumulant expansion over the qubit operators, which matches numerical calculations to great accuracy, was devised in Ref.\,\cite{FelixPRA}. In Ref.\,\cite{FUR} we have further provided the relevant expressions $\langle J_y(\tau)\rangle, \langle J_y^2(\tau)\rangle$ for non-collective spin-boson dephasing, where qubit $n$ is assumed to be at position $\vec{r}_n$; analogous formulas for $\langle J_x(\tau)\rangle, \langle J_x^2(\tau)\rangle$ may be readily deduced. Here, we work in 1D, with $r_n \equiv n x_0$ in the limit of short interrogation time $\omega_c \tau \ll 1$. In this regime, the quantum noise (phase) contributions to the dynamics may be disregarded to first approximation, as we argued in the main text. The relevant cumulant expansion for the required expectation values can then be greatly simplified:
\begin{align}
 \langle J_x(\tau) \rangle &\approx  e^{- \kappa_0^2 (\omega_c \tau)^2} \mathcal{Q}_0(\theta, \beta) \cos(\varphi ) ,  \label{exp0}\\ 
 \langle J_y(\tau) \rangle & \approx  e^{- \kappa_0^2 (\omega_c \tau)^2} \mathcal{Q}_0(\theta, \beta)\sin(\varphi ), 
   \\
    \langle J_v^2(\tau)\rangle &\approx \frac{N}{4} + e^{-2 \kappa_0^2 (\omega_c \tau)^2}\left[C_1(\theta, \beta)\, G_{+}( x_0) \pm \cos( 2 \varphi) \, C_2(\theta, \beta) \, G_{-}( x_0) \right] ,\quad v \in \{x,y\},
\label{exp}
\end{align}
with $G_{\pm}( x_0)\equiv \sum_{j=1}^{N-1}\, (N-j)\, e^{ \pm \,2 \kappa_{0}^2( \omega_c\tau)^2 \tilde{g}_1(j x_0)} $ capturing the lattice spatial correlations, and $\mathcal{Q}_0(\theta, \beta), C_1(\theta, \beta)$, $C_2(\theta, \beta) $ being functions of the angles $\theta, \beta$. In the $N\gg 1$ limit we are interested in, the latter take the form:
\begin{align}
    Q_0(\theta,\beta)&=\frac{N}{2} e^{N \theta^2/8 }, 
    \label{Q} \\
    C_1(\theta, \beta)&=\frac{1}{8} \,\left[ 2 \,\cos(\beta)^2 - (1+e^{-N \theta^2/2}) \sin(\beta)^2 - 2 e^{-N \theta^2/2} \sin(2 \beta)\sin\left( \theta/2\right) \right] ,\\
    C_2(\theta, \beta)&= \frac{1}{16} \left[2 \sin(\beta )^2 + 4 \sin (2 \beta ) \sin \left(\theta/2\right) e^{-N \theta ^2/8}+( \cos (2 \beta ) + 3 )\, e^{-N \theta ^2/2 }\right].
\end{align}

We emphasize that numerical calculations show that, in the time regime of interest, for as low as $N=30$ qubits, the ratio uncertainty which follows from these approximate expressions already closely reproduces its counterpart {\em including} quantum noise contributions, whose expressions are much more involved; see inset in Fig.\,\ref{both}(right) of the main text. 
  
\subsubsection{Optimal noise-tailored precision scaling}
\label{optOAT}

Let us now derive the optimal sensor performance for an OATS under {\em classical} dephasing (that is, ignoring contributions from the quantum spectra, hence $\varphi_0(t)$). As remarked in the main text, the squeezing and twisting angles were chosen to minimize the initial variance along the $y$ axis, according to Eq.\,\eqref{ideal}. The numerator and denominator in $\Delta \hat{b}^{\text{OAT}}_R (\tau)$, obtained by substituting the OATS mean values from Eqs.\,(\ref{exp0})-(\ref{exp}) into Eq.\,(\ref{dbR}) must first be minimized with respect to their phase argument $\varphi$. Since an optimally squeezed initial variance along $y$ has a maximally anti-squeezed component along $x$, the ideal phase is such that this latter term vanishes: $\varphi= k \pi, k \in \mathbb{N} $. In this case, Eq.\,(\ref{dbR}) becomes {\em proportional} to the precision derived from method of moments when measuring the observable $\mathcal{O}= J_y$: 
\begin{equation}
\Delta \hat{b}_{\text{R}}^{\text{OAT}}(\tau)^2 =\frac{ \Delta J_y^2(\tau)}{ T \tau \, \langle J_x (\tau)\rangle^2}, 
\label{oatR}
\end{equation}
where we used $\partial_{\varphi} \langle J_y (\tau)\rangle= \langle J_x (\tau)\rangle$. For the same fixed total time $2T$, however, when we are {\em only} measuring $J_y$ and computing the precision through the method of moments, we have $\Delta \hat{b}^{\text{OAT}}(\tau)^2= \Delta J_y^2(\tau)/ (2 T \tau  \,[\partial_{\varphi} \langle J_y (\tau)\rangle]^2\,)$: that is, the variance of the ratio estimator is larger by a factor 2.

The variance $\Delta J_y^2(\tau)$ in the numerator of Eq.\,\eqref{oatR} can now be expanded in powers of $(\omega_c \tau)$ up to the fourth order, whereas we only need to keep the denominators' leading-order contribution; replacing in $\Delta \hat{b}_R (\tau)$ yields  
\begin{eqnarray}
 \Delta \hat{b}_{\text{R}}(\tau) \approx  \frac{ \left[\,a_0(N,x_0) + a_2(N,x_0)(\omega_c \tau)^2 + a_4(N,x_0)\,(\omega_c \tau)^4 \, \right]^{1/2}}{\sqrt{T \tau}\, h_0(N) }, \label{dbST}
 \end{eqnarray}
where the coefficients have a well-defined $N$-dependence through the ideal squeezing and twisting angles of Eq.\,(\ref{ideal}), as well as a spatial dependence coming from the $G_{\pm}(x_0)$ in the short-time expansion. Analytic formulas for the coefficients can be derived by means of the Polygamma functions approach worked out for the GHZ state in Sec.\,\ref{GHZstate}. This yields $h_0(N) \approx N/2$ and 
\begin{eqnarray*}
a_0(N,x_0)& \approx & \frac{3^{2/3}}{8} N^{1/3} ,\\
a_2(N,x_0)& \approx & \frac{1}{2} \Gamma(s+1) \bigg\{ \frac{1}{ 6 x_0^2} + 4 \bigg(\frac{\pi }{x_0}\bigg)^4 e^{-\frac{2 \pi }{x_0}} \bigg[ \bigg(  \frac{1}{3}\bigg)^{1/3} N^{1/3} - \bigg(\frac{1}{3}\bigg)^{2/3} N^{2/3}+\frac{1}{3} N\bigg] 
  +   \frac{1}{2} \big( 3^{1/3} N^{2/3}-3^{2/3} N^{1/3} \big) \bigg\} ,  \\
a_4(N,x_0)&\approx & \bigg\{\left[ \Gamma (s+3) \left( \frac{1}{16\,3^{1/3}} N^{1/3} - \frac{1 }{16 \, 3^{2/3}} N^{2/3} \right) +\frac{1}{2}\, N\, \Gamma (s+1)^2\right]-\frac{\Gamma (s+3)}{480} \frac{1}{x_0^2}  \\ 
& \mbox{} & + \,\, \Gamma (s+3) e^{-\frac{2 \pi }{x_0}} \left(\frac{\pi }{x_0}\right)^6 \left( -N^{1/3} \frac{1}{30\,3^{1/3}}+\frac{1}{30 \, 3^{2/3}} N^{2/3}+ N\frac{1}{90}\right)  \bigg\} .
\end{eqnarray*}

Similar to the GHZ state, Eq.\,(\ref{dbST}) allows us to minimize with respect to $\tau$ and $x_0$ independently. We can  obtain the optimal interrogation time by setting the first derivative  of Eq.\,(\ref{dbST}) with respect to $\tau$ equal to zero:
\begin{eqnarray}
\tau^{\text{OATS}}_{\text{R  opt}} = \frac{1}{\sqrt{6}  a_4(N,x_0)} \left[ \left(12 a_0(N,x_0) a_4(N,x_0)+a_2(N,x_0)^2\right)^{1/2} -a_2(N,x_0)  \right]^{1/2} 
\label{toptOATS}.
\end{eqnarray}
Evaluating Eq.\,(\ref{dbST}) at $\tau^{\text{OATS}}_{\text{R  opt}}$ leads, in turn, to the following approximate time-optimized frequency uncertainty:
\begin{eqnarray}
\Delta \hat{b}_{\text{R}}(\tau^{\text{OATS}}_{\text{R opt}}) &=&\frac{2^{5/4} }{3^{3/4}} \, \frac{\,[a_2(x_0,N) \Delta+12 a_0(N,x_0) a_4(N,x_0)]^{1/2}}{\, h_0(N) N \left[ a_4(N,x_0) \Delta\right]^{1/4}} ,
\label{dboatsopt} \\
\Delta &\equiv& \left(12 a_0(N,x_0) a_4(N,x_0)+a_2(N,x_0)^2\right)^{1/2}\!\!-a_2(N,x_0). \notag
\end{eqnarray}
Analytic minimization with respect to $x_0$, however, is involved. Numerical evidence shows that the value of the ideal lattice parameter is remarkably close to the value $x_{0 \text{ an}}^{\text{OATS}}$ that minimizes $a_2(x_0,N)$. Following a similar procedure as in deriving Eq.\,(\ref{vopt}), we can then provide an asymptotic solution to the equation $\partial_{x_0} a_2(x_0,N) =0$:
\begin{eqnarray*}
x_{0 \text{ an}}^{\text{OATS}}= - \frac{2 \pi}{3} \log \left[ \frac{\log (3 \pi^{2/3} N^{1/3} ) }{3 \pi^{2/3} N^{1/3}} \right].
\end{eqnarray*} 

\begin{figure}[t!]
    \centering
    \includegraphics[width=7cm]{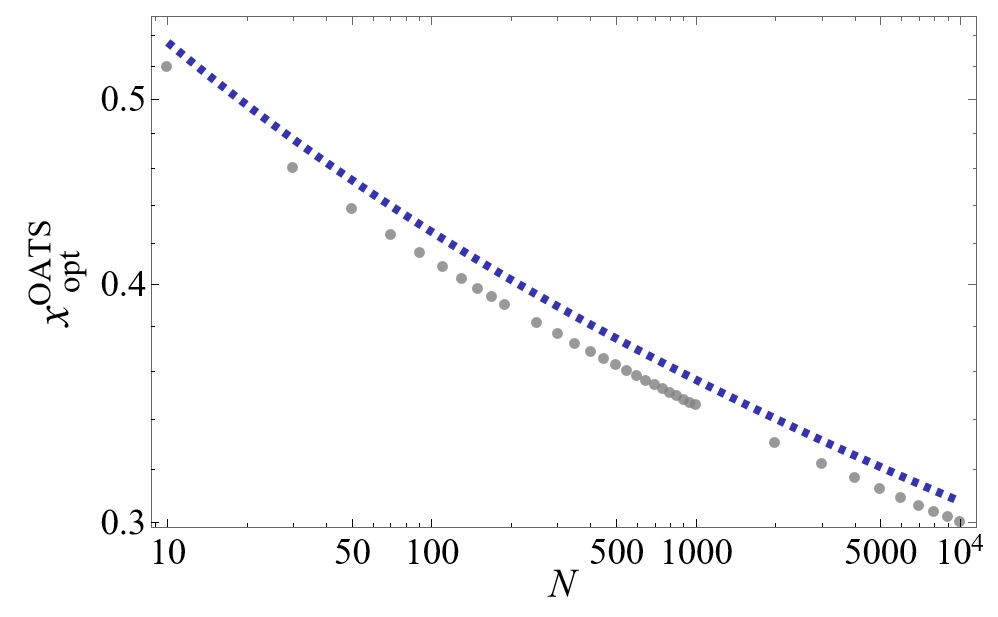}
    \vspace*{-2mm}
    \caption{{\bf Optimal lattice separation for OAT states under classical noise. }
Grey dots: Exact numerical optimization, yielding $x_{\text{ opt}}^{\text{OATS}}$. Blue dashed line: Approximate analytic solution $x_{\text{opt an}}^{\text{OATS}}$, Eq.\,(\ref{xopt}). A supra-Ohmic spectral density with an exponential cutoff, and $\alpha=1, s=3,  \omega_c=1$ is assumed. }
    \label{xoptplot}
\end{figure}

\noindent 
Much like for the GHZ state, the values of $x_{0 \text{ an}}^{\text{OATS}}$ differ from the numerical optimization by a small offset, see Fig.\,\ref{xoptplot}; nonetheless, parameter robustness leads to the analytic value obtained by computing $\Delta \hat{b}_{\text{R}}(\tau^{\text{OATS}}_{\text{R  opt}}, x_{0 \text{ an}}^{\text{OATS}} )$ to be extremely close to the exact minimum. Evaluating the expansion coefficients at $x_{0 \text{ an}}^{\text{OATS}}$, and disregarding all non-leading orders in $N$, we get:
\begin{eqnarray*}
a_0(N,x_{\text{ an}}^{\text{OATS}}) \approx \frac{3^{2/3}}{ 2^{7/3}} N^{1/3} ,\;\;\;\;a_2(N,x_{\text{ an}}^{\text{OATS}}) \approx \frac{3^{1/3}}{2^{5/3}}\, \Gamma (s+1)\,N^{2/3},\;\;\;\;\;a_4(N,x_{\text{an}}^{\text{OATS}}) \approx \frac{1}{4}\;\Gamma (s+1)^2\,N.
\end{eqnarray*}
For supra-Ohmic spectral densities with $s=3$, in particular, Zeno-like scaling is then found upon substituting the above approximate expressions in Eq.\,(\ref{dboatsopt}),
\begin{eqnarray}
\Delta \hat{b}_{\text{R opt}}^{\text{OATS}}(N) \approx \frac{1}{\sqrt{3}} \,\Gamma (s+1)^{1/4} \sqrt{\frac{\omega_c}{T}} N^{-3/4},  
\label{zeno}
\end{eqnarray}
attained at an optimal time $\tau_{\text{R opt}}^{\text{OATS}} \approx 3^{-1/3} \,\Gamma(s+1)^{-1/2}  N^{-1/6}$, in excellent agreement with numerical optimization. We remark that, based on numerical fittings, Eq.\,(\ref{zeno}) has been found to hold for {\em all $s>1$ spectral densities} we investigated, although it has only been rigorously derived for $s=3$.

\vfill 

\twocolumngrid


%

\end{document}